\documentstyle[12pt,epsfig]{article}

\newcommand{\be}{\begin{equation}}
\newcommand{\ee}{\end{equation}}
\def\aprle{\buildrel < \over {_{\sim}}}
\def\aprge{\buildrel > \over {_{\sim}}}
\begin{document}
\topmargin 0pt
\oddsidemargin=-0.4truecm
\evensidemargin=-0.4truecm
\renewcommand{\thefootnote}{\fnsymbol{footnote}}
\newpage
\setcounter{page}{1}
\begin{titlepage}     
\vspace*{-2.0cm}
\begin{flushright}
IC/98/98 \\
\vspace*{-0.2cm}
hep-ph/9808270
\end{flushright}
\vspace*{-0.5cm}
\begin{center}
{\Large \bf Atmospheric
neutrinos at  Super-Kamiokande and parametric resonance in  neutrino
oscillations}
\vspace{0.5cm}

{E. Kh. Akhmedov$^1$
\footnote{On leave from National Research Centre Kurchatov Institute, 
Moscow 123182, Russia. E-mail: akhmedov@sissa.it}}, 
A. Dighe$^1$ 
\footnote{E-mail: amol@ictp.trieste.it}, 
P. Lipari$^2$ , 
A. Yu. Smirnov$^1$
\footnote{Also at the Institute of Nuclear Research of Russian Academy 
of Sciences, Moscow 117312, Russia. E-mail: smirnov@ictp.trieste.it}\\
{\em (1) The Abdus Salam International Centre for Theoretical Physics,  
I-34100 Trieste, Italy }
{\em (2) University of Rome, ``La Sapienza" and INFN, 
I-00185 Rome, Italy }

\end{center}
\vglue 0.8truecm
\begin{abstract}
We consider the oscillations of atmospheric neutrinos 
in the earth in the three-neutrino scheme with a
$\Delta m^2$ hierarchy and a small  
admixture of the electron neutrino in the
heavy mass eigenstate characterized by the mixing angle 
$\theta_{13}$. We show that for $\Delta m^2 \simeq (0.5 - 3)\times 10^{-3}$ 
eV$^2$ indicated by the Super-Kamiokande data and 
$\sin^2 2\theta_{13} \aprle 0.2$, 
the oscillations of multi-GeV neutrinos 
in the subdominant $\nu_{\mu} \leftrightarrow \nu_e$
mode are enhanced by the MSW and parametric resonances.  
The parametric resonance, which occurs when the neutrinos 
cross the core of the earth, dominates for  
$\Delta m^2 \simeq (1 - 2)\times 10^{-3}$ eV$^2$, 
$\sin^2 2\theta_{13} \aprle 0.06$.  
The resonance matter effects 
lead to an observable  excess of the e-like events with a specific 
zenith angle dependence even for small $\theta_{13}$. The 
up-down asymmetry of the multi-GeV e-like events 
can reach 15\% for $|\cos \Theta_e| > 0.2 $
and up to 30\% for $|\cos \Theta_e| > 0.6 $, 
where $\Theta_e$ is the zenith angle of the electron. 
The resonance matter effects are relevant for the interpretation 
of the Super-Kamiokande data.

\end{abstract}
\end{titlepage}
\renewcommand{\thefootnote}{\arabic{footnote}}
\setcounter{footnote}{0}
\newpage
\section{Introduction}
Neutrino oscillations in matter can be strongly enhanced if the matter
density varies periodically  along the neutrino path.
In this case the parametric resonance of neutrino oscillations can occur
\cite{ETC,Akh}. The probability of the  transition of a neutrino  from one
flavor state to another may become close to unity even 
when the mixing angles (both in vacuum and in matter) are small. 
The parametric effects are further enhanced if the parametric resonance
energy is close to that of the MSW resonance \cite{KS}.  

The simplest realization of the periodic matter density distribution, 
which in addition is of practical importance, is the periodic 
step function (``castle wall'') profile. In this case,
the density modulation period $L$ 
consists of two parts, $L_c$ and $L_m$, which correspond to constant but
different matter densities $N_c$ and $N_m$. The evolution equation for
neutrino oscillations in matter with such a density distribution    
allows for an exact analytic solution \cite{Akh}. 

For the ``castle wall'' density profile, the parametric resonance 
conditions are especially simple: the lowest-order (principal) resonance 
occurs when the oscillation phases $\phi_m$ and $\phi_c$ defined through 
$\phi_i=2\pi L_i/l_m(N_i)$ [($i=m,c$) with $l_m(N_i)$ being the oscillation 
length in the matter with density $N_i$]
satisfy \cite{Akh,LS}
\be
\phi_c = \phi_m = \pi\,.
\label{param1}
\ee
   
Recently it has been pointed out \cite{LS} that
atmospheric
neutrinos traversing the earth pass through layers of alternating
density and can therefore undergo parametrically enhanced oscillations.
Indeed, the earth consists of two main structures -- the mantle
and the core. The matter density changes rather slowly within the mantle 
and within the core but  at their border it jumps sharply by about a 
factor of two. Therefore  to a good approximation, one may consider 
the mantle and the core as structures of constant densities equal to the 
corresponding average densities (two-layer model) 
\footnote{A comparison of neutrino oscillation 
probabilities calculated with such a simplified matter density profile
with those calculated with actual density profile provided by geophysical
models shows a very good agreement (for recent discussions see, 
{\it e.g.}, \cite{LS,LiLu}).}.
Neutrinos coming to the detector from the lower hemisphere at zenith
angles $\Theta_\nu$ in the range defined by 
$\cos \Theta_\nu=(-1) \div (-0.837)$  
traverse the earth's mantle, core and then
again mantle. Therefore such neutrinos experience a periodic 
``castle wall'' potential. 
Even though the neutrinos pass only through ``one and a half'' periods 
of density modulations (this would be exactly one and a half 
periods if the distances neutrinos travel in the mantle and core were
equal), the parametric effects on neutrino oscillations in the earth can be 
quite strong \cite{LS,P,Akh1}. 

In \cite{LS} the effect of parametric resonance  on possible 
$\nu_{\mu} \leftrightarrow \nu_{sterile}$ 
oscillations of atmospheric neutrinos was
considered. The effect was found to be potentially important for zenith angle 
distributions of through-going and stopping muons produced by high energy 
neutrinos. 
The parametric enhancement of atmospheric neutrino oscillations  
$\nu_{\mu} \leftrightarrow \nu_e$
was discovered numerically in \cite{numer1}; however, in these papers the
parametric nature of the enhancement was not recognized and possible 
consequences for atmospheric neutrino oscillations were not fully studied. 

The parametric enhancement can also take place for oscillations of 
solar neutrinos in the earth. It was realized in \cite{P,Akh1} 
that
a sizeable enhancement of the regeneration effect for neutrinos 
crossing the core and having the energies lying between the MSW resonance
energies in the core and in the mantle, found in \cite{numer2},
 is due to the 
parametric resonance.

In the present paper we discuss in detail possible manifestations of the 
resonance matter effects and in particular, of the parametric resonance 
in {\em flavor} oscillations of atmospheric neutrinos.
The Super-Kamiokande (SK) collaboration has recently reported a strong
evidence for neutrino oscillations in their atmospheric neutrino data 
\cite{SuperK}, confirming the previously observed 
anomaly in the flavor composition of atmospheric neutrinos \cite{ANA}. 

The SK group has analyzed their data in the framework of two-flavor 
oscillations. They have shown that 
the data can be fitted well assuming 
$\nu_\mu\leftrightarrow \nu_\tau$ oscillations with the maximal or
close to maximal mixing ($\sin^2 2\theta \simeq 1$) and 
$\Delta m^2=(0.5 - 6)\times 10^{-3}$ eV$^2$.  
Pure $\nu_{\mu} \leftrightarrow \nu_e$ oscillations are practically
excluded: they would have resulted in a significant zenith angle dependence 
of the e-like events in contradiction with the observations. In addition, 
the sizeable deficiency of muon-like events would require a strong
(close to maximal) $\nu_e-\nu_\mu$ mixing, which is excluded by 
the CHOOZ experiment \cite{CHOOZ} for most of the values of 
$\Delta m^2$ relevant for the atmospheric neutrino anomaly. 
At the same time, $\nu_{\mu} 
\leftrightarrow \nu_e$ oscillations with small mixing angles are still 
possible. Moreover, the data shows some excess of e-like events 
both in sub-GeV and multi-GeV samples and  
therefore is suggestive of the $\nu_{\mu} \leftrightarrow \nu_e$ 
oscillations as a subdominant mode. In the data analysis by the SK
collaboration the excess of e-like events was accounted for by up-scaling
the overall normalization of the fluxes of atmospheric neutrinos by 
factors which are compatible with the uncertainty of the theoretical 
predictions. However, recent cosmic ray measurements by the BESS experiment 
\cite{BESS} indicate that the overall normalization of the atmospheric 
neutrino fluxes 
in the existing theoretical predictions was rather 
overestimated than underestimated. 
This makes the problem of the excess of  e-like events more serious. 
In addition, the data, though not yet conclusive, seems to indicate 
some deviation of the zenith-angle dependence of the e-like events 
from the dependence that
follows just from the zenith-angle dependence of the original flux
in the absence of oscillations. 
Furthermore,  it is difficult to explain the low value of the 
double ratio $R = (e/\mu)/(e/\mu)_{MC}$ 
in a wide range of neutrino energies 
just in terms of the $\nu_{\mu} \leftrightarrow \nu_{\tau}$  oscillations 
\cite{LoSecco}. 
All this may indicate that the electron neutrinos are also involved 
in the oscillations.

In this paper we study the atmospheric neutrino oscillations in the 
3-flavor scheme with $\nu_\mu\leftrightarrow \nu_\tau$ 
being the dominant channel. We assume that the vacuum $\nu_\mu-\nu_e$
mixing angle satisfies the upper bounds following from
the CHOOZ data. We show that the MSW and parametric enhancements of
atmospheric neutrino oscillations occur in the subdominant 
$\nu_\mu \leftrightarrow \nu_e$ mode, leading to observable effects despite 
the smallness of the mixing angle. 

There have been a number of studies of the atmospheric neutrino oscillations 
with the three  neutrino mixing
\cite{threenu1,threenu2,threenu3,threenu4}. 
However, their authors have either concentrated on gross 
characteristics of the atmospheric neutrino oscillations (such as  
the allowed values of mixing angles and mass squared differences) 
\cite{threenu1}, or considered only the cases when all mixing angles
are large and/or $\Delta m_{atm}^2$ is large \cite{threenu2}, or  
employed the constant-density approximation of the structure of the earth
\cite{threenu3}, or neglected the matter effects on neutrino oscillations 
in the earth \cite{threenu4}.  
In the first case the results are rather insensitive to the parametric
effects, while in the last three cases these effects are missed altogether. 

Although the MSW enhancement effects are important and for a wide
range of parameters dominate the excess of the multi-GeV e-like events, 
we concentrate here on the parametric resonance effects for two reasons: 
(1) in contrast to the MSW enhancement which was widely discussed in the
past the parametric effects in the oscillations of atmospheric neutrinos 
have not been studied in detail; (2) the parametric resonance modifies 
the MSW resonance peaks.  

The paper is organized as follows. 
In Sec. 2 we find the probabilities of oscillations 
in the three neutrino system in terms of the $\nu_e - \tilde{\nu}_3$ 
oscillation probability which experiences the resonance enhancement. 
In Sec. 3 we consider the parametric resonance effects on 
the atmospheric neutrinos. In Sec. 4 we present 
the results of the numerical calculations for the zenith angle dependence  
and the up-down asymmetry of the e-like and $\mu-$like events and confront
the results with 
observations. In Sec. 5 we discuss our results as well as the prospects 
of observing the parametric resonance in the oscillations of atmospheric
neutrinos. The details of the calculations of the cross-sections and the
event rates are given in the Appendix. 

\section{Three-flavor oscillations of atmospheric neutrinos}

We consider the three-flavor neutrino system  with a hierarchy of mass 
squared differences 
\be
\Delta m^2_{21} \ll \Delta m^2_{32} \approx \Delta m^2_{31} ~~.
\label{hier}
\ee
We assume that $\Delta m_{32}^2 \equiv \Delta m_{atm}^2 \aprge
5\times 10^{-4}$
eV$^2$ is relevant for the atmospheric neutrino oscillations whereas 
$\Delta m_{21}^2 \aprle 10^{-5}$ eV$^2$ allows one to solve the  solar
neutrino problem either through the MSW effect or through the long range 
vacuum oscillations.

The evolution of the neutrino vector of state $\nu_f \equiv (\nu_e,
\nu_{\mu}, \nu_{\tau})^T$ is described by the equation 
\be
i \frac{d \nu_f}{dt} =
\left( \frac{U M^2 U^\dagger}{2 E} + V \right) \nu_f, 
\label{evolution}
\ee 
where $E$ is the neutrino energy and  $M^2 = diag (m^2_1, m_2^2, m_3^2)$
is the diagonal matrix of neutrino mass squared eigenvalues. 
$V = diag(V_e, 0 ,0)$ is the matrix of matter-induced neutrino potentials
with $V_e = \sqrt 2 G_F N_e$, $G_F$ and $N_e$ being the Fermi constant 
and the electron number density, respectively.   
The mixing matrix $U$, defined through $\nu_f = U \nu_{m}$ where $\nu_{m}=
(\nu_1, \nu_2, \nu_3)^T$ is the vector of neutrino mass eigenstates, can 
be parametrized as 
\be
U = U_{23} U_{13} U_{12}.   
\label{mixing}
\ee
The matrices $U_{ij}= U_{ij}(\theta_{ij})$ perform the rotation  
in the $ij$- plane by the angle $\theta_{ij}$. 
We have neglected possible CP-violation effects in the lepton sector 
which are strongly suppressed in the case of the mass hierarchy
(\ref{hier}).  

Let us introduce new states 
$\tilde{\nu} = (\nu_e, \tilde{\nu}_{2}, \tilde{\nu}_{3})^T$  
obtained by performing the $U_{23}$ - transformation: 
$\nu_f = U_{23} \tilde{\nu}$. The Hamiltonian $\tilde{H}$ that 
describes the evolution of the $\tilde{\nu}$ state can be obtained from 
(\ref{evolution}) and (\ref{mixing}): 
$$
\tilde{H} =  \frac{1}{2E} U_{13} U_{12} M^2
U^\dagger_{12} U^\dagger_{13} ~ + ~ V ~.
$$ 
We get  explicitly 
\be
\tilde{H} \approx  \left(\begin{array}{ccc}
s_{13}^2 \Delta m_{32}^2/2E + V_e   & 0
& s_{13}c_{13} \Delta m_{32}^2/2E  \\
0 & c_{12}^2 \Delta m_{21}^2/2E  &   0    \\
 s_{13}c_{13} \Delta m_{32}^2/2E
&  0  & c_{13}^2 \Delta m_{32}^2/2E
\end{array}\right)\, , 
\label{matr1}
\ee
($c_{13} \equiv \cos \theta_{13}$, $s_{13} \equiv \sin \theta_{13}$, 
etc.)   after  the following approximations.  
Since  $\Delta m_{32}^2$ is in the range  $\aprge 5\times 10^{-4}$
eV$^2$ 
while $m_{21}^2 < 10^{-5} {\rm eV}^2$,  
the  terms of the order $s_{12}^2 \Delta m_{21}^2/ \Delta m_{32}^2$ 
were neglected. 
Also, $s_{12}c_{12}\Delta m_{21}^2/2EV_e \aprle 10^{-3}$, so  
the  (12)-element in the matrix (\ref{matr1}) 
(i.e. the mixing between the $\nu_e$ and $\tilde{\nu}_2$) was 
neglected.

According to (\ref{matr1}),  the $\tilde{\nu}_2$ state decouples from the
rest of the system
and evolves independently. Therefore the S-matrix (the matrix of  
amplitudes) in the basis $(\nu_e, \tilde{\nu}_2, \tilde{\nu}_3)$  
has the following form :
\be  
S \approx 
\left(\begin{array}{ccc} 
A_{ee}   & 0      & A_{e3} \\
0        & A_{22} &   0    \\
A_{3e}  & 0 & A_{33} 
\end{array}
\right) ~~ ,~~
\label{matr2}
\ee
where 
\be 
A_{22} = \exp(-i\phi_{2})\,, \quad \quad      
\phi_{2} = \frac{c_{12}^2 \Delta m_{21}^2 L}{2E},  
\label{phase}
\ee
and  $L$ is the total distance traveled by the neutrinos. 
Notice that in our approximation  $\phi_{2}$ 
does not depend on the matter density.  
The $(\nu_e, \tilde{\nu}_3)$ subsystem 
evolves according to the  2$\times$2 Hamiltonian 
[$\nu_e - \tilde{\nu}_3$ submatrix in (\ref{matr1})] 
determined by the potential $V_e$, mixing
angle $\theta_{13}$ and 
the mass squared difference $\Delta m^2_{32}$. 
Let us denote by  
\be
P_2 \equiv |A_{e3}|^2 = |A_{3e}|^2 =    
1 - |A_{ee}|^2 = 1 - |A_{33}|^2
\ee 
the probability of the  $\nu_e \leftrightarrow
\tilde{\nu}_{3}$ oscillations.  As we will show in Sec. 3,  
it is in this channel that the oscillations 
are parametrically enhanced. 

The $S$-matrix in the flavor basis can be obtained from 
(\ref{matr2}) by $U_{23}$ rotation: $U_{23} S U_{23}^{\dagger}$. 
Then 
the probabilities  of flavor oscillations in the three neutrino system  
can  be found as 
$P(\nu_\alpha \to \nu_\beta)=|(U_{23} S U_{23}^{\dagger}) _{\alpha \beta}|^2$,  
which yields 
\be
P(\nu_e\to \nu_e) = 1 - P_2 ~~, 
\label{ee}
\ee
\be
P(\nu_e\to \nu_\mu) = P(\nu_\mu\to \nu_e) = s_{23}^2  P_2 ~~, 
\ee
\be
P(\nu_e\to \nu_\tau) = c_{23}^2 P_2 ~~, 
\ee
\be
P(\nu_\mu\to\nu_\mu ) = 1  - s_{23}^4 P_2 +
2s_{23}^2 c_{23}^2\,\left[Re (e^{-i\phi_2} A_{33}) - 1 \right]~~, 
\ee
\be
P(\nu_\mu\to\nu_\tau ) = s_{23}^2\,c_{23}^2\,
\left[ 2 - P_2 - 2Re (e^{-i\phi_2} A_{33}) \right]. 
\label{mutau}
\ee  
The phase  $\phi_2$ is defined in (\ref{phase}). 
The interpretation of the above results is straightforward.  
For instance, the $\nu_{\mu} \leftrightarrow \nu_e$ 
transition occurs via the projection of $\nu_{\mu}$ onto  
$\tilde{\nu}_{3}$ and $2\nu$ - oscillations 
$\tilde{\nu}_{3} \leftrightarrow \nu_e$. 
The $\nu_{\mu} \leftrightarrow \nu_{\tau}$ 
transition can occur in two different ways: 
(i) $\nu_{\mu}$ projects onto $\tilde{\nu}_{2}$, the latter propagates
without transition and at the final time of evolution is  projected
onto $\nu_{\tau}$; (ii) $\nu_{\mu}$ projects onto $\tilde{\nu}_{3}$;  
since $\tilde{\nu}_{3}$ oscillates into $\nu_e$, at the final time one 
should project the amplitude of the survival probability of 
$\tilde{\nu}_{3}$ onto $\nu_{\tau}$. The interference of these 
two transition amplitudes leads to the probability (\ref{mutau}).   

Using the probabilities (\ref{ee}) - (\ref{mutau}) one can find the 
modifications of the atmospheric neutrino fluxes due to the oscillations. 
Let  $F_e^0$ and $F_{\mu}^0$ be the electron and muon neutrino fluxes at 
the detector in the absence of  oscillations. Then the fluxes in the
presence of  oscillations can be written as 
\be
F_e = F_e^0 \left[ 1 + P_2(r s_{23}^2 - 1)\right]~~, 
\label{fluxe}
\ee
\be 
F_{\mu} = F_{\mu}^0 \left[ 1 - s_{23}^4 
\left(1 - \frac{1}{r s_{23}^2} \right)
P_2   
+ 2s_{23}^2 c_{23}^2 \left[ Re (e^{-i\phi_2} A_{33}) - 1 \right] \right], 
\label{fluxmu}
\ee
where 
$$
r(E, \Theta_\nu) = \frac{F_{\mu}^0(E, \Theta_\nu)}{ F_e^0(E, \Theta_\nu)}
$$ 
is the ratio of the original muon and electron neutrino fluxes.

It is interesting that one can have either an excess or a deficiency of
e-like events depending on the values of $r$ and
$s_{23}$. Indeed, the effect of oscillations on the electron neutrino flux
is proportional to the factor $(r s_{23}^2 - 1)$. If one assumes 
$r = 2$, there will be an excess of e-like events for 
$\theta_{23}>45^\circ$ and a deficiency for $\theta_{23}<45^\circ$. The 
SK best fit was $\theta_{23}=45^\circ$; in this case there would be 
no deviation from the prediction for $r=2$. However, for upward going 
neutrinos in the multi-GeV range $r$ is typically 3 -- 3.5 rather than 2, 
so there should be an excess of e-like events even if $\theta_{23}=45^\circ$. 
In addition, notice that 
the SK analyses were performed in the two-flavor scheme, 
and the best-fit value of $\theta_{23}$ may be somewhat different in the 
3-flavor analysis. 

A final remark is that in the two neutrino mixing
scenario, the probability of 
$\nu_{\mu} \leftrightarrow \nu_{\tau}$
oscillations depends on $\sin^2 2\theta_{23}$, and hence has an ambiguity 
$\theta_{23} \leftrightarrow (\pi/2 - \theta_{23})$. In the three neutrino 
case the probability $P(\nu_e\leftrightarrow \nu_\mu)$ depends 
on $s^2_{23}$, so that the study of an excess of the e-like events allows 
one to resolve the ambiguity.  
 

\section{Parametric enhancement of neutrino oscillations}

The amplitude $A_{33}$ and the probability $P_2$ which enter into the 
expressions (\ref{ee}) - (\ref{mutau}) have to be found by solving the
evolution equation for the $(\nu_e, \tilde{\nu}_3)$ system. The
transitions in this system are the ones that undergo the
resonance (parametric and MSW) enhancements.  One can study properties 
of the resonance matter effects using the two-layer model 
of the earth's density profile.   
In the two-layer model  $P_2$ and $A_{33}$ 
can be found explicitly in a compact form \cite{Akh1}:  
\be
P_2 = \left(2\sin \frac{\phi_m}{2} \sin 2\theta_m \,Y + 
\sin \frac{\phi_c}{2} \sin 2\theta_c \right)^2\,, 
\label{prob}
\ee
\begin{eqnarray} 
Re [e^{-i\phi_2} A_{33}] = \left(2 \cos \frac{\phi_m}{2} \,Y - 
\cos \frac{\phi_c}{2}\right)
\cos (\Phi-\phi_2) - \nonumber \\
\left(2 \sin \frac{\phi_m}{2} \cos 2 \theta_m \,Y + \sin \frac{\phi_c}{2}
\cos 
2\theta_c \right)\sin(\Phi-\phi_2)\,, 
\end{eqnarray} 
where 
\be
Y \equiv  \cos \frac{\phi_m}{2} \cos \frac{\phi_c}{2} - 
\sin \frac{\phi_m}{2} \sin \frac{\phi_c}{2}  
\cos (2 \theta_m -  2 \theta_c)\,.  
\ee 
Here $\phi_m$ and $\phi_c$ are the oscillation phases 
acquired by the 
neutrino system in the mantle (one layer) and in the core, respectively. 
They can be written as 
\be
\phi_i = \frac{2 \pi L_i}{l_m({V_i})} = 
{L_i}\, \Delta H (V_i)~,~~~~ (i = m, c),   
\label{phases}
\ee
where 
\be
\Delta H(V_i) =  
\sqrt{\left(\cos 2\theta_{13} \frac{\Delta m^2_{32}}{2E} -V_i\right)^2 
+ \left(\sin 2\theta_{13} \frac{\Delta m^2_{32}}{2E} \right)^2  }
\ee
is the level splitting (difference between  the eigenvalues 
of $H$), $V_m$ and
$V_c$
being the potentials in the mantle and in the core. The angles 
$\theta_m$ and  $\theta_c$ are the values of the 
$\nu_e - \tilde{\nu}_3$ 
mixing angle  in matter of the mantle and the core 
respectively. They  can be found from  
\be
\sin 2\theta_i = \sin 2\theta_{13} \, \frac{\Delta m^2_{32}}
{2E\Delta H(V_i)}\,,~~~ (i = m, c)\,. 
\ee
The phase $\Phi$ is given by the integral of $(\Delta m_{32}^2/4E +V_e/2)$
along the 
neutrino path between its production and detection points. In the
two-layer model of the earth's density profile,  it is given by 
\be
\Phi= \left(\frac{\Delta m^2_{32}}{2E}  + V_m \right) ~ L_m + 
 \left(\frac{\Delta m^2_{32}}{2E}  + V_c \right) ~ \frac{L_c}{2} 
\ee
(we neglect the neutrino oscillations in the air which are unimportant 
for the range of the neutrino parameters of interest). 
For neutrinos crossing both the core and  the mantle 
($\sin^2 \Theta_\nu < (R_c/R_\oplus)^2 = 0.299$) 
the path lengths in the mantle and the core are
determined by the relations 
\be
L_m = R_\oplus \, \left(-\cos\Theta_\nu - \sqrt{(R_c/R_\oplus)^2 - 
\sin^2\Theta_\nu}\,\right)\,,   
\label{Lm}
\ee
\be
L_c = 2R_\oplus \,\sqrt{(R_c/R_\oplus)^2 - \sin^2\Theta_\nu}\  . 
\label{Lc}
\ee
Here $R_\oplus$ is the radius of the earth and $R_c$ is the radius 
of the core. For $\sin^2 \Theta_\nu\ge 0.299$ neutrinos cross 
the mantle only and 
their path length is $2 L_m = -2 R_\oplus \,\cos\Theta_\nu$.   
 
The physical picture of the oscillations 
and the resulting event rates   
depend crucially on the neutrino parameters $\Delta m_{32}^2$
and $\sin^2 2\theta_{13}$ as well as on the zenith angle 
$\Theta_\nu$. Let us consider the dependence of the oscillation 
probability on the neutrino energy. In the region 
\be
\frac{E}{\Delta m^2_{32}} = (0.8  - 4)\cdot 10^{12}~~ {\rm eV}^{-1} 
\label{region}
\ee
neutrinos experience resonantly enhanced oscillations in matter. 
This interval is determined by the MSW resonance energies for 
oscillations in the core and in the mantle $E_c^R$ and $E_m^R$   
\be
E_i^R = \frac{\Delta m_{32}^2 \cos 2 \theta_{13}}{ 2 V_i} ~~~~~(i = c, m) 
\label{Eres}
\ee
and by the resonance widths $\Delta E/ E_i^R \sim 2\tan  2\theta_{13}$.  
The MSW resonance enhancement leads to characteristic peaks in the 
energy dependence of the transition probability. 
The exact positions of the maxima of the peaks depend on the oscillation
phases and in general do not coincide with the resonance energies
(\ref{Eres}). Neutrinos having the  trajectories with $\cos\Theta_\nu >
-0.84$ do not cross the core of the earth and therefore for such neutrinos  
only the MSW resonance enhancement of the oscillations in the mantle
can occur. 

For $\cos\Theta_\nu < - 0.84$ there is an interference between the 
oscillation effects in the core and in the mantle which strongly depends
on $\sin 2\theta_{13}$.  
For $\sin^2 2\theta_{13} > 0.15$ there is a significant overlap of the 
MSW resonances in the core and in the mantle. The interference 
leads to a rather complicated picture of neutrino oscillations in the
overlap region with a modification of the MSW resonance peaks.  
The parametric resonance conditions are not fulfilled. 

For $\sin^2 2\theta_{13} < 0.15$ the probability $P_2$ as a function 
of $E$ has three main peaks: 
two peaks with the maxima at $\sim E_c^R$ 
and $\sim E_m^R$ due to the MSW resonance oscillations in the core   
and the mantle respectively, and a peak between them (fig. 1a). The 
latter is due to the parametric enhancement of oscillations. 
In what follows we shall call for brevity the resonance peaks due to 
the MSW effects for neutrinos oscillations in the earth' mantle and core 
{\it the mantle peak} and {\it the core peak} respectively. 

The maximum of the parametric peak is at an energy $E_p$ 
at which the resonance conditions (\ref{param1}) are satisfied; 
the analysis \cite{P,Akh1} of the parametric resonance condition 
shows that a significant parametric enhancement occurs only when 
\be
\frac{E}{\Delta m^2_{32}} \simeq (1  - 2)\cdot 10^{12}~ {\rm eV}^{-1}\,, 
\label{region1}
\ee 
{\it i.e.} $E_p$ is indeed in the range $E_c^R <  E_p < E_m^R$. 

At the maximum  of the parametric peak the transition probability 
(\ref{prob}) takes the value \cite{LS} 
\be 
P_2^{max} = \sin^2 2(\theta_c - 2\theta_m)\, 
\label{prob7}
\ee
provided that the resonance conditions are exactly fulfilled. 
Due to a small number of periods (``1.5 period"),   
the energy width of the parametric resonance $\Delta E/E_p$ is  
large. The transition probability decreases
by a factor of two for \cite{Akh1} 
\be
\Delta\phi_{i}=|\phi_{i}-\pi|\simeq \frac{\pi}{2}\,,~~(i = c, m). 
\label{detun}
\ee
Thus, the resonance enhancement of neutrino oscillations can occur 
even for quite sizeable detuning of the phases $\phi_{m,c}$. 
Numerically we get $\Delta E/E_p \sim 2 - 3$. 

Let us stress that the interference of the mantle and the core 
oscillation effects leads not only to the appearance of the parametric
peak. It also modifies significantly the MSW peaks as compared with 
the peaks which would appear in the one layer cases without interference.
In fact, the interference leads to a suppression of those peaks. 

The parametric resonance conditions (\ref{param1}) constrain the allowed
values of $\theta_{13}$. 
If these conditions are to be satisfied for all the neutrino
trajectories that cross the core (including the vertical ones) one gets from 
(\ref{param1}) the upper limit $\sin^2 2\theta_{13}\le \pi^2/(4(L_c)_{max}^2
\,V_c^2) \simeq 0.04$ \cite{Akh1}. 
If the zenith angles close to $180^\circ$ are  excluded, the 
constraint becomes less stringent. For example, for $\sin^2 \Theta_\nu \ge
0.12$ one obtains $\sin^2 2\theta_{13}\le 0.07$.  

The main features of the parametric peak are illustrated in 
fig. 1. 
For $\cos \Theta_{\nu} \simeq  - 1$ the parametric 
peak and the core peak partially overlap (actually the core peak
appears as a shoulder on the low-energy slope of the parametric peak). 
With increasing $\cos \Theta_{\nu}$ the parametric peak moves towards larger
energies; it becomes well resolved from the MSW peaks in the interval 
$\cos\Theta_{\nu} = (-0.94)\div (-0.87)$ and eventually transforms 
into the mantle peak at $\cos \Theta_{\nu} > -0.85$. 

The relative strength (area) of the parametric and the MSW peaks depends
on the value of $\theta_{13}$. As follows from fig. 1b, with decreasing
$\theta_{13}$ the MSW peaks decrease faster than the parametric peak. 
This can be explained as follows. For a given value of $\theta_{13}$ 
and $\Theta_\nu$ there is in general some  detuning at the parametric 
peak, i.e. the conditions (\ref{param1}) are only approximately
satisfied. With decreasing $\sin^2 2\theta_{13}$ 
the maximal possible value (\ref{prob7}) of the probability $P_2$,  
which corresponds to the exact parametric resonance, decreases.  
At the same time, it turns out that
the detuning  becomes smaller which partially compensates  
the decrease in  $P_2^{max}$.  Therefore in some range of
$\theta_{13}$ the 
parametric peak lowers only moderately with decreasing $\sin^2 2\theta_{13}$. 
In contrast to this,  the MSW peaks decrease rather quickly with 
$\sin^2 2\theta_{13}$. 
As a result, for  $\sin^2 2\theta_{13} < 0.06$  
the parametric peak has the largest strength. 

To summarize, the parametric peak is the most pronounced in the ranges of 
the parameters characterized by eq. (\ref{region1}) and 
\be 
\sin^2 2\theta_{13} = (1 - 12)\cdot 10^{-2}\,,  ~~~~ 
\cos \Theta_{\nu} = (- 1) \div (-0.84)\,. 
\ee
For the  sub-GeV sample of e-like and $\mu$-like events the relevant
energies of neutrinos are $E \approx 0.3 - 1.5$ GeV. Then from 
(\ref{region1}) we find that  for the sub-GeV events 
a significant effect of the parametric resonance 
is expected if the mass squared difference  is in the range  
\be
\Delta m^2_{32} \approx (1 - 10)\cdot 10^{-4}~ {\rm eV}^2~.  
\label{range1}
\ee 
The multi-GeV sample gets its main contribution from neutrinos with
energies  
$E \approx 1.3 - 10$ GeV and the corresponding $\Delta m^2_{32}$ are
larger: 
\be
\Delta m^2_{32} = (0.4 - 4)\cdot 10^{-3}~ {\rm eV}^2.
\label{range2}
\ee
Quite interestingly, the central value of this range  coincides 
with the best fit value of $\Delta m^2_{32}$ which follows from the  
analysis of the contained events in SK \cite{SuperK}. 

The parametric resonance can play an  important role  
for certain samples of events ({\it e.g.} multi-GeV)  
which pick up a relatively narrow neutrino energy  interval. 
The number of events is determined by the integral  over neutrino
energy $E$  
of the oscillation probability folded in with the response function 
$f(E)$:  $N \propto \int d E f(E) P (E)$. 
The response function  describes the contribution of 
neutrinos with energy $E$ to a given sample of events. 
In particular, the response function for the multi-GeV events 
in the Super-Kamiokande  has a maximum at $E\approx 2.5$ GeV. 
The value of $f$ increases rapidly with energy below the 
maximum but has a rather long tail above it; 
the energies at the half-height are $E\simeq 1.5$ and 4 GeV.
Thus the width of the response function is characterized by a factor 
of 2 - 3. 

As can be seen from fig. 1, for $\Delta m^2_{32} \simeq (1. 5 - 2)
\times 10^{-3}~{\rm eV}^2$ and values of $\cos\Theta_\nu$ 
varying in the range $(-1) \div (-0.84)$ 
(i.e. covering the earth's core), the parametric peak spans
essentially the whole region of energies that corresponds to the peak 
of the response function of multi-GeV neutrinos. 
effect in the core). The mantle peak is in the 
tail of the response function and therefore its contribution to the 
excess of e-like events is attenuated. 
For $\sin^2 2\theta_{13} < 0.06$ the contribution of the parametric peak
dominates for trajectories crossing the core of the earth.   
However, even for higher values of $\theta_{13}$ ($\sin^2 
2\theta_{13} \aprle 0.15$) the contribution of the parametric resonance 
to the excess of e-like events can be comparable to that of the 
mantle and core peaks provided 
that $\Delta m^2_{32} \sim (1 - 2) \times 10^{-3}~{\rm eV}^2$.  

For $\Delta m^2_{32} \aprle 10^{-3}~{\rm eV}^2$ the overlap 
of the parametric peak and the peak of the response function is small 
and the main contribution to the excess of e-like events comes from the 
mantle peak. 

For $\Delta m^2_{32} > (2 - 3) \times 10^{-3}~{\rm eV}^2$, 
with increasing  $\Delta m^2_{32}$ the resonance effects weaken and
the oscillation effects are essentially reduced to those of vacuum
oscillations. 


\section{Resonance effects and zenith angle  dependence of  
e-like and $\mu$-like events}


The resonance matter effects in the atmospheric neutrino oscillations 
can manifest themselves as an enhanced excess of the e-like
events with a specific zenith angle dependence. 
We calculate  the zenith angle distributions of the 
e-like and $\mu$-like events for 
$\Delta m^2_{32} = (0.1 - 10)\times 10^{-3}~{\rm eV}^2$, 
$\sin^2 2\theta_{23} \aprge 0.7$ indicated by the SK data  
and a wide range of values of $\sin^2 2 \theta_{13}$ 
taking the CHOOZ bounds into account. 
We examine whether the matter effects,
and in particular the parametric resonance, can be relevant for
understanding such features of the SK data as the excess of the e-like 
events and the asymmetries of e-like and $\mu$-like events. 

The number of e-like or $\mu$-like events with the detected charged 
lepton in the energy interval $\Delta E_l$  and direction $\Omega_l$  
$(l= e, \mu)$ can be calculated as
\begin{eqnarray}
N_l (\Omega_l) & = &  \sum_{\nu, \bar{\nu}}
 \int_{\Delta E_l} dE_{l} \,
 \int d\Omega_{\nu l} \,
 \int d\Omega_{\nu} \,
 \int dE_{\nu} \,
F_l(E_{\nu}, \Omega_\nu) \times  \nonumber \\
& & ~~~~~~~~~~~
\frac{d^2 \sigma (E_{\nu}, E_l, \Omega_{\nu l})}{d E_l \, d\Omega_{\nu l}
}\,
\, \delta [\Omega_l -  (\Omega_{\nu} \oplus \Omega_{\nu l})]\, \epsilon
(E_l)~,
\label{int}
\end{eqnarray}
where $F_l(E_{\nu}, \Omega_{\nu})$, $(l = e, \mu)$ are the fluxes of
neutrinos in the detector defined in eqs. (\ref{fluxe}), (\ref{fluxmu});
$E_l$ is the energy of the charged lepton.
Up to the small geomagnetic effects the neutrino fluxes and
therefore the charged lepton distributions depend only on the zenith angle:
$N_l (\cos \Theta_l) = 2 \pi N_l(\Omega_l)$.  
$\Omega_{\nu l}$ is the angle between the directions of
the incoming neutrino and the outgoing lepton,
$d^2 \sigma/ d E_l d\Omega_{\nu l} $ is the neutrino charged current
cross-section, $\epsilon (E_l)$ is the charged lepton detection efficiency,  
and  the integration goes over $E_l > 1.3$ GeV ($E_l< 1.3$ GeV) for 
multi-GeV (sub-GeV) sample. 

In calculating the neutrino fluxes at the detector we have used the 
two-layer model of the earth's structure [eqs. (\ref{prob}) - (\ref{Lc})]; 
the average densities of the core and the mantle were calculated for each
neutrino trajectory using the actual density profile provided by the Stacey 
model \cite{Stacey}. For $\Delta m_{21}^2<10^{-5}$ eV$^2$ the correction
due to the phase $\phi_2$ in eq. (\ref{fluxmu}) is very small, and we have 
put $\phi_2=0$ in the actual calculations. The details of the calculations 
of the cross-sections and the decay rates are described in the Appendix. 

The integration over the neutrino zenith angles and energies 
leads to a significant smearing of the 
$\Theta_l$ dependence. Indeed,  
the average angle between the neutrino and the outgoing charged lepton  is
about $15^{\circ} - 20^{\circ}$ in the multi-GeV region and it is almost
$60^{\circ}$ in the sub-GeV range. 
Therefore  the data do not give us  direct
information about the zenith angle dependence of $\nu_e$ and $\nu_\mu$ 
fluxes at the detector. In particular, a significant contribution to the
vertical upward bin $\cos\Theta_l=(-1\div -0.8)$ comes from the
neutrinos which cross 
the mantle only and therefore do not experience the parametric enhancement
of oscillations. Thus, the observed parametric enhancement
effects are weakened as compared to the enhancement in the neutrino zenith 
angle distributions. This is especially true for the sub-GeV sample.  
Additional smearing of the neutrino zenith angle dependence due to 
the finite 
angular and energy resolution is relatively small. 

In the antineutrino channels,  matter suppresses oscillations 
\footnote{We assume $\Delta m_{32}^2 >0$ throughout the paper.} 
and consequently the parametric effects are weak. 
Since neutrinos and antineutrinos of a given flavor are not distinguished 
in the atmospheric neutrino experiments 
({\it i.e.} only total $\nu_e+\bar{\nu}_e$ 
and $\nu_\mu+\bar{\nu}_\mu$ fluxes are measured), the resonance enhancement 
effects are  additionally diluted.  

Let us first consider the multi-GeV events. In Fig.~2  we show the zenith
angle dependences of the e-like and $\mu$-like events for a representative 
set of the oscillation parameters $\Delta m^2_{32}$, $\sin^2 2\theta_{13}$ 
and $\sin^2 \theta_{23}$. The e-like events exhibit an excess which first 
appears in the horizontal bin ($\cos \Theta_e  \simeq 0$) and increases 
monotonically with decreasing $\cos \Theta_e$. It reaches $\sim 20$\% in 
the vertical (upward) bin, $\cos \Theta_e \simeq -1$. 
In contrast to this, $\mu$-like events exhibit a deficiency at 
$\cos\Theta_\mu < 0$.  

A convenient quantitative measure of the excess or deficiency of
e-like and $\mu$-like events in the vertical bins is the up-down asymmetry   
\be
A^{U/D}_{l}(b_1,b_2)  = 2~ \frac
{N_{l}^{up}(b_1,b_2) - N_{l}^{down}(b_1,b_2)}
{N_{l}^{up}(b_1,b_2) + N_{l}^{down}(b_1,b_2)}~, ~~~~~ (l = e,\,\mu)~, 
\ee
where 
\be
N_{l}^{up}(b_1,b_2) = \int^{-b_1}_{-b_2} d \cos \Theta_l
N_{l}(\Theta_l)~,~~~ 
N_{l}^{down}(b_1,b_2) = \int^{b_2}_{b_1} d \cos\Theta_l N_{l}(\Theta_l)~, 
\ee
and $N_{l}(\cos\Theta_l)$ are  given  in (\ref{int}). 
The parametric as well as the MSW resonance enhancement effects are
largest 
in the vertical (upward) bins. 
In  most of our calculations 
we use $b_1 = 0.6$, $b_2=1$ which corresponds to    
the SK binning. Notice that about 60\% of neutrinos contributing to
this bin have the trajectories which cross the mantle only and therefore 
do not undergo the parametric enhancement of oscillations. 

{}From the  SK results \cite{SuperK} we find  
\be 
A^{U/D}_e (0.6,1) \approx 0.22 \pm 0.21 ~.
\label{asymm}
\ee 
The up-down asymmetry and the excess of the e-like events increase 
with increasing $s_{23}^2$. In order to assess the maximal possible 
effects we therefore choose for most of our calculation 
the value $s_{23}^2=0.75$ which is close to the upper border of the
values allowed by the SK data. 

In fig.~3a  we show  the dependence 
of the up-down asymmetry $A^{U/D}_e(0.6,1)$ on 
$\Delta m^2_{32}$ for  
$s^2_{23}= 0.75$ and  different values of $\sin^2 2\theta_{13}$.  
The dependence  of $A^{U/D}_e$ on $\Delta m^2_{32}$ reflects the changing 
degree of overlap of the response function with the  parametric peak as well 
as the MSW peaks (see Sec. 3).  
The maximum of the asymmetry 
($A_e^{U/D}(0.6,1) \simeq 0.15 - 0.25$) is  at 
$\Delta m^2_{32} \approx (0.8 - 2) \times 10^{-3}$ eV$^2$. 
Thus, our calculations for $\sin^2 2\theta_{13} \aprge 0.06$ reproduce 
the central value of the up-down asymmetry observed by the SK (\ref{asymm}). 
They also reproduce (within $1\sigma$) the experimentally observed
excess of e-like events in the vertical bin. 

With decreasing $\sin^2 2\theta_{13}$ the asymmetry decreases and 
the maximum shifts to larger values of $\Delta m^2$. Indeed, for large 
$\theta_{13}$ ( $\sin^2 2\theta_{13} \sim  0.1$) the mantle peak 
gives the main contribution and the maximum of asymmetry at 
$\Delta m^2_{32} =  0.8  \times 10^{-3}$ eV$^2$ reflects the position 
of this peak. With decreasing $\theta_{13}$ the parametric peak
becomes relatively more important and the maximum of asymmetry 
shifts to $\Delta m^2_{32} \approx (1.5  - 1.7) \times 10^{-3}$ eV$^2$ 
at $\sin^2 2\theta_{13} \sim  0.06$, which 
corresponds to the position of the parametric peak. With  
further  decrease of $\theta_{13}$ the position of the maximum remains 
unchanged, in accordance with fig.~1b.  
We find that for $\sin^2 2\theta_{13} \sim  0.025$ the contribution
of the parametric peak to the excess of the e-like events is about 60\%. 

With  decreasing $\Delta m^2_{32}$ the asymmetry decreases 
since the mixing in matter (and consequently the oscillation
probability) decreases. For $\Delta m^2_{32} < 0.3 \times 10^{-3}$ 
eV$^2$ it approaches the asymmetry due to the geomagnetic effect without 
oscillations. 

For $\Delta m^2_{32} > 3\times  10^{-3}$ eV$^2$ the asymmetry 
decreases with increasing $\Delta m^2_{32}$ for two reasons. 
(i) The matter enhancement of mixing disappears and the oscillation effect 
is essentially reduced to that of {\it vacuum} oscillations 
governed by the small $\sin^2 2\theta_{13}$ 
(the oscillations due to  the matter splitting between
the levels of the two light eigenstates are suppressed by the
factor $\sin^2 2 \theta_{13} \cdot \sin^2 2 \theta_{23}/4$
\cite{threenu3}); 
(ii)  for $\Delta m_{32}^2\aprge 5\times 10^{-2}$ eV$^2$ the
oscillations become  important for down-going neutrinos too and the 
up-down asymmetry goes to zero.

In fig.~3b we show the asymmetry of $\mu$-like events,  
which is opposite in sign  to that for the e-like events. 
The absolute value of the asymmetry becomes maximal at 
$\Delta m^2_{32} \sim 0.3 \times 10^{-3}$ eV$^2$. 
This corresponds to the situation when 
the average distance in the vertical bin ($\sim 10^4$ km) 
equals half of the vacuum oscillation 
length for a typical energy of multi-GeV neutrinos $E\sim 2$ GeV. 
The matter effects are important in the range  
$\Delta m^2_{32} \sim (0.5 - 5) \times 10^{-3}$ eV$^2$ where 
the absolute value of the asymmetry increases with $\sin^2 2\theta_{13}$
just as in  fig. 3a. 
We show by dotted line the asymmetry which would be expected in the case 
of pure $\nu_{\mu} \leftrightarrow \nu_{\tau}$ oscillations with the same 
$s_{23}^2 = 0.75$.  
(For $2\times 10^{-3}\aprle \Delta m_{32}^2/{\rm eV}^2\aprle 10^{-2}$ 
it can be  roughly estimated assuming that in this range the oscillations 
of upward going neutrinos are fully averaged  whereas the downward going 
neutrinos do not oscillate at all, which yields  
$A_{\mu}^{U/D} \approx 
\sin^2 2\theta_{23}/(2 - 0.5 \sin^2 2\theta_{23})  = - 0.46$.) 
The difference between the dotted line and the others 
shows the enhancement of the asymmetry due to  an  
additional channel of oscillations 
$\nu_{\mu} \leftrightarrow \nu_{e}$ in the three neutrino system. 
In the  case of  $\nu_{\mu} \leftrightarrow \nu_{\tau}$ oscillations 
with maximal mixing the asymmetry is larger:  
$A_{\mu}^{U/D} \sim - 0.67$.  Both values are compatible with the 
SK result, $-0.56 \pm 0.15$ \cite{SuperK}. 

As we have pointed out above, the parametric enhancement of
neutrino oscillations occurs for neutrinos crossing
the earth's core ($\cos \Theta_\nu < -0.84$). 
Therefore the largest up-down asymmetry is achieved when the
binning enhances the contribution of the core-crossing neutrinos. 
In fig.~3c we compare the electron asymmetries in the bins 
$0.84\le |\cos \Theta_{\nu_e}| \le 1$, 
$0.60\le|\cos \Theta_{\nu_e}| \le 0.84$ 
and  $0.60\le |\cos \Theta_{\nu_e}| \le 1$. 
One can see that the asymmetry is largest for the first bin which 
has the maximal contribution from the neutrinos 
crossing the core of the earth.  
The maximum of asymmetry is achieved at 
$\Delta m_{32}^2\simeq (1.5 - 1.7)\times 10^{-3}$ eV$^2$ which is typical of 
the parametric resonance at multi-GeV energies. The position of the peak
of the asymmetry in the second bin which is dominated by the mantle-only
crossing neutrinos ($\Delta m_{32}^2 \simeq 0.7\times 10^{-3}$ eV$^2$) 
reflects the position of the MSW peak in 
the mantle. The parametric peak is higher than the MSW one because for   
small $\sin^2 2\theta_{13}$ the parametric effects dominate. 
In the third (largest) bin, $0.60\le|\cos \Theta_{\nu_e}| \le 0.84$, 
the parametric effects are significantly diluted compared to the first 
bin. 

To summarize, the physical picture of the neutrino oscillations
depends crucially on the value of $\Delta m^2_{32}$. 
The whole range of $\Delta m^2_{32}$ can be 
divided into  three parts:\\ 
(1) the region of the vacuum oscillations: 
$\Delta m^2_{32} \aprge 3 \times  10^{-3}$ eV$^2$  
for multi-GeV neutrinos (here there is also a 
small effect due to the matter-induced level splitting of 
the light states);\\  
(2) the resonance region (two MSW resonances and the parametric
resonance):  
$\Delta m^2_{32} \simeq (0.5 - 3) \times 10^{-3}$ eV$^2$;\\ 
(3) the region of matter suppressed 
oscillations $\Delta m^2_{32} \aprle 0.5 \times 10^{-3}$ eV$^2$.

For sub-GeV neutrinos the corresponding regions are shifted 
by about a factor of 3 to smaller values of $\Delta m^2_{32}$. 

In fig.~4a, we show the dependence of the up-down asymmetry 
$A^{U/D}(0.6,1)$ on $\sin^2 2\theta_{13}$ 
for  $s^2_{23} = 0.75$ and several values of $\Delta m^2_{32}$.  
The dashed curve corresponds to the value 
$\Delta m^2_{32}= 1.7 \times 10^{-3}$  eV$^2$ from the resonance region. 
The asymmetry (and the excess)  of the e-like events rapidly
increases with $\sin^2 2\theta_{13}$ in the region  $\sin^2 2\theta_{13} 
\aprle 0.10$ which corresponds to the parametric resonance, 
and then increases more slowly. 

The  solid line corresponds to a value of $\Delta m^2_{32}$ close to 
the vacuum oscillation region. For $\sin^2 2\theta_{13} > 0.08$ 
the asymmetry increases almost linearly  with $\sin^2 2\theta_{13}$,  
as is expected in the case of vacuum oscillations.   
Here, too, the CHOOZ bound is important: 
for the allowed values ($\sin^2 2\theta_{13} < 0.16$) the 
asymmetry is smaller than 
that for the value $\Delta m^2_{32}=1.7\times 10^{-3}$ eV$^2$ 
from  the resonance region. 
The dot-dashed line represents the asymmetry in the 
region of the  matter suppressed oscillations: 
the effective mixing is suppressed even for large $\sin^2 2\theta_{13}$ 
and the asymmetry is relatively small. In fig. 4b, we show the
corresponding asymmetry in the $\mu$-like events. 

As follows from (\ref{fluxe}), the asymmetry in the e-like events 
is proportional to the factor $(\bar{r} s^2_{23} - 1)$, where
$\bar{r} \sim 2.5$ is the ratio of the original muon and electron neutrino
fluxes averaged over the zenith angles and energies in the multi-GeV 
sample.  According to this, the asymmetry due to the oscillations increases 
almost linearly with $s^2_{23}$ (fig. 5a); it becomes zero at
$s^2_{23}\sim \bar{r}^{-1} \sim 0.4$, where the total asymmetry,  
$A \sim 0.05$, equals  the small asymmetry due to the geomagnetic effects. 

In fig.~5b, we show the dependence of the  up-down asymmetry for 
$\mu$-like events on $s^2_{23}$. For pure  
$\nu_{\mu} \leftrightarrow  \nu_{\tau}$ vacuum 
oscillations (dotted curve), the up-down asymmetry is a symmetric 
function with respect to 
$s^2_{23} \leftrightarrow (1 - s^2_{23})$. 
Matter effects  
break this symmetry and the breaking  increases with 
$s^2_{23}$ in accordance with our previous analysis.  

With increasing $s^2_{23}$ the excess of the e-like 
events gets enhanced.  
At the same time, for $\theta_{23}>45^\circ$, 
the probability of the $\nu_\mu\leftrightarrow \nu_\tau$ 
oscillations decreases with increasing $s_{23}^2$. The additional channel 
of oscillations $\nu_{\mu} \leftrightarrow \nu_e$ does 
not compensate for this decrease, and so the asymmetry and the 
total suppression of the $\mu$ - like events become smaller 
(see fig.~5 a, b).

The zenith angle dependence of the multi-GeV e-like events 
with and without $\nu_e \leftrightarrow \nu_{\mu}$ oscillations 
is shown in fig.~6, along with the SK data for 535 days. 
One can see that taking into account the $\nu_e\leftrightarrow 
\nu_\mu$ oscillations improves the fit of the data but cannot fully 
explain the excess of the e-like events unless the overall normalization 
of the atmospheric $\nu_e$ and $\nu_\mu$ fluxes is increased. 

In fig.~7 the iso-asymmetry curves for multi-GeV e-like events are  
plotted in the $(\sin^2 2\theta_{13}$, $\Delta m_{32}^2)$ plane, 
along with the constraints from the CHOOZ experiment (shaded area). 
The behavior of the curves 
can be understood from the preceding discussion. 
The large-$\Delta m_{32}^2$ region of the plot corresponds to the vacuum
oscillations; vertical and near-vertical lines are due to the averaged 
oscillations with no or little dependence of probability on $\Delta
m_{32}^2$. 
Matter effects increase the asymmetry in the region 
$\Delta m_{32}^2\simeq (0.5 - 3) \times 10^{-3}$ eV$^2$: 
the iso-asymmetry curves are ``pulled'' towards the region of small 
values of $\sin^2 2\theta_{13}$. 
For $\Delta m_{32}^2\aprle 0.5 \times 10^{-3}$ eV$^2$ matter suppresses 
the $\nu_e\leftrightarrow \nu_\mu$ oscillations. 

The parameter space of the resonance region can be divided into three
parts:
(i) $\Delta m_{32}^2\aprle  10^{-3}$ eV$^2$, where the mantle resonance 
dominates; 
(ii) $\Delta m_{32}^2 \aprge 1.5 \times 10^{-3}$ eV$^2$, 
$\sin^2 2\theta_{13} <  0.06$, where the parametric resonance 
gives an important contribution;   
(iii) $\Delta m_{32}^2 \aprge 1.5 \times 10^{-3}$ eV$^2$, 
$\sin^2 2\theta_{13}\simeq 0.06 - 0.12$, where  parametric peak gives a 
smaller contribution which, however, is comparable to that due to the
mantle resonance. The core resonance gives a smaller effect.

For $\sin^2 2\theta_{13} >  0.15$ there is a complex interference 
of the MSW resonance effects in the mantle and core. 

The CHOOZ bound 
excludes the part of the parameter space that corresponds to large 
asymmetry, $A^{U/D}_e(0.6,1)>0.28$. There is a local maximum of the 
asymmetry $A^{U/D}_e(0.6,1)\simeq 0.27$ at 
$\sin^2 2\theta_{13}\simeq 0.14$, $\Delta m_{32}^2\simeq 10^{-3}$ eV$^2$. 
Values close to this can also be achieved at the same  
$\Delta m_{32}^2\simeq  10^{-3}$ eV$^2$ but large mixing angles
$\sin^2 2\theta_{13}\simeq 0.5 - 0.9$. 
{}From fig.~7 it follows that 
the asymmetry $A^{U/D}_e(0.6,1)$ can be as large as 0.22 in the range 
of parameters (ii), i.e. rather close to the maximal possible value 
indicated above.  

The parametric enhancement of the $\nu_e\leftrightarrow \nu_\mu$ 
oscillations makes it possible to have a sizeable asymmetry even for 
very small values of the mixing angle $\theta_{13}$: the asymmetry can 
be as large as 0.15 even for $\sin^2 2\theta_{13}=0.02$ provided that 
$\Delta m_{32}^2$ lies in the range $(1 - 2)\times 10^{-3}$ eV$^2$.  
Notice that in the absence of matter effects one would expect 
the asymmetry to be an order of magnitude smaller. 

The iso-asymmetry plot of fig.~7 has been obtained for the fixed value 
of $s_{23}^2$ ($s_{23}^2=0.75)$; the magnitudes of the asymmetry for
other values of $s_{23}^2$ can be easily found using the relation 
$A_e^{U/D} = 2x/(2+x)$, where $x=(s_{23}^2\bar{r}-1)P_2$. 

As we have pointed out above, the largest asymmetry allowed by the
CHOOZ constraints, $A_e^{U/D}(0.6,1)\simeq 0.28$, is achieved in a small 
region around $\Delta m_{32}^2 \simeq 0.8\times 10^{-3}$ eV$^2$ and
$\sin^2 2\theta_{13}\simeq 1$ (see fig.~7). 
Future reactor experiments, and in particular KAMLAND 
\cite{KAMLAND}, will be able to probe this range of parameters. 
If no oscillations are found, the values 
$\sin^2 2\theta_{13}\ge 0.1$ will be excluded. 
In this case the largest possible asymmetry would correspond to the 
small-$\theta_{13}$ region where the parametric resonance effects 
play an important role. It should also be emphasized that in case 
of the negative result of KAMLAND 
the studies of the excess of the e-like events in atmospheric neutrinos 
would be a unique way to probe the parameter range $\sin^2 2\theta_{13} 
<0.1$, $\Delta m_{32}^2>5\times 10^{-4}$ eV$^2$. 

The calculated event rates are sensitive to the response function 
used. The latter depends on the event selection criteria, 
detection efficiency  and other features of detector, {\it etc.}.
The results given above correspond to the response function 
described in Sec. 3. 
The response function with the maximum (and median energy) 
shifted by about 25\% to higher energies compared to what 
we used would shift the iso-asymmetry contours by about 
25\% to larger $\Delta m^2$. Notice that in this case the region 
of large asymmetries at large mixing angles will be completely 
excluded and the largest asymmetry would be achieved at 
$\sin^2 2\theta_{13}\sim 0.15$.  

Let us now consider possible effects in the sub-GeV sample. 
In fig.~8 we show the zenith angle dependence of the sub-GeV events  
for $\Delta m^2_{32} = 0.3 \times 10^{-3}$ eV$^2$ which 
corresponds to the maximum of the resonance effects.   
Due to the strong smearing, the excess has a rather weak 
zenith angle dependence. Moreover, even in the vertical bin 
it does not exceed 10\%. The up-down asymmetry is smaller than 0.03.  
Notice that in the sub-GeV region  
$\bar{r}$ is closer to 2 than in the multi-GeV region  and the factor 
$\bar{r} s_{23}^2-1$ in (\ref{fluxe}) 
leads to an additional suppression of the transition probability.
 
In  fig.~9a we show the dependence of ratio $N_e /N_e^0$ of the
total numbers of the sub-GeV events on $\Delta m^2_{32}$   
with and without oscillations. 
The matter oscillations enhance the excess of e-like events in the range 
$\Delta m^2_{32} = (0.1 - 1 ) \times 10^{-3}$ eV$^2$    
and the maximum is at $\Delta m^2_{32} \simeq 0.3 \times 10^{-3}$ eV$^2$.

For   $\Delta m^2_{32} > 10^{-3}$ eV$^2$, 
which corresponds to the maximal effect in the multi-GeV sample,  
the oscillation effect in the sub-GeV sample approaches that of the 
vacuum oscillations.
For $\sin^2 2 \theta_{13} < 0.1$, we  
find that the total excess of the e-like events 
is below 5\% and the asymmetry is  below 3\%. 

The ratio $N_{\mu} /N_{\mu}^0$ is mainly determined by the $\nu_\mu
\rightarrow 
\nu_\tau$ oscillations. In the 2-flavor case ($\theta_{13}=0$) these
oscillations are unaffected by matter. Therefore in the 3-flavor case 
for small values of $\theta_{13}$ the matter effects 
on $N_{\mu} /N_{\mu}^0$ are relatively small (fig.~9b).


\section{Discussion and conclusions}


The excess of e-like events both in multi-GeV and sub-GeV samples 
observed by the SK experiment may indicate that the electron neutrinos
are involved  in the oscillations of atmospheric neutrinos. 
In this connection we have considered the  oscillations of atmospheric 
neutrinos in the $3\nu$ scheme. 
Assuming the mass hierarchy $\Delta m_{32}^2 \approx \Delta m_{31}^2 \gg 
\Delta m_{21}^2$ we have derived simple analytic expressions for the 
oscillation probabilities in the $3\nu$ system in terms of the oscillation 
amplitudes in the $2\nu$ system $(\nu_e, \tilde{\nu}_3)$, where 
$\tilde{\nu}_3$ is a linear combination of $\nu_\mu$ and $\nu_\tau$. 
For the amplitudes of the $\nu_e\leftrightarrow \tilde{\nu}_3$
oscillations in the earth analytic expressions obtained in the two-layer
model of the earth's structure were used. 

Let us summarize our main results. 
We have shown that the range of the neutrino parameters 
$\sin^2 2\theta_{13}\aprle 0.2$, $\Delta m_{32}^2\approx (0.5 - 3)\times
10^{-3}$ eV$^2$ 
is the resonance range for multi-GeV events in which the oscillations in
the subdominant $\nu_e\leftrightarrow \tilde{\nu}_3$ mode are strongly 
enhanced by matter effects. 

For $\sin^2 2\theta_{13} <  0.1$ and neutrino trajectories 
crossing the core of the earth the transition probability  
$\nu_{\mu} \leftrightarrow \nu_e$  as the function of energy  
has three peaks:  two MSW peaks due to the resonance enhancement of
the oscillations in the mantle and in the core and the parametric
resonance peak between them. The parametric peak   
dominates over the MSW peaks for $\sin^2 2 \theta_{13} < 0.06$. 
 
For $\sin^2 2\theta_{13} > 0.15$ the energy intervals for  
the MSW resonances in the core and mantle strongly overlap and 
complex interference phenomena occur.  

The resonance effects manifest themselves  in the zenith angle dependences 
of the  charged leptons produced in the interactions of neutrinos, although
the integration over the neutrino angles and energies leads to a smearing 
of the zenith angle distribution of charged leptons. 

We have found that the asymmetry $A^{U/D}_e(0.6,1)$ of the multi-GeV
e-like events can be as large as about 0.28 for the domain of parameters  
allowed by the CHOOZ bound and $s_{23}^2=0.75$.    
This value is achieved at $\Delta m_{32}^2\simeq 10^{-3}$ eV$^2$ and   
$\sin^2 2\theta_{13}\aprge 0.6$. The parametric resonance
is not operative at such large values of $\sin^2 2\theta_{13}$.  

The asymmetry in the parameter region 
$\sin^2 2\theta_{13}\aprle 0.06$, 
$\Delta m_{32}^2\approx (1 - 2)\times 10^{-3}$ eV$^2$
where the parametric resonance becomes important 
(contributes $\aprge 60\%$) 
can be as large as  
$A^{U/D}_e(0.6,1) \simeq 0.22$, i.e. is close to the maximal 
possible value. 

An important consequence of the parametric effects is that even for very 
small values of the mixing angle $\theta_{13}$ quite a sizeable asymmetry
of the multi-GeV e-like events can result. The asymmetry can be as large 
as 0.15 even for $\sin^2 2\theta_{13}\simeq 0.02$. 

The oscillations of the sub-GeV 
neutrinos could be parametrically enhanced only for about a factor of 3 
smaller values of $\Delta m_{32}^2$, close to the lower bound of the range 
allowed by the SK data. However even in this case the parametric resonance 
would not affect the asymmetries of the sub-GeV data significantly. 

We have found that taking into account the subdominant $\nu_e\leftrightarrow 
\nu_\mu$ oscillations leads to an excess of e-likes events and improves
the fit of both multi-GeV and sub-GeV e-like data in the SK experiment. 
However, for all allowed values of the oscillation parameters the predicted 
excess ($\sim 3 - 5\%$) is smaller than the observed one. 
Thus, if the observed excess survives future experimental tests, one will 
need alternative explanations for it.  

The following remarks are in order. 

1. If the excess of multi-GeV e-like 
events is at least partly due to the $\nu_e\leftrightarrow \nu_\mu$ 
oscillations, it leads to a lower bound on the mixing angle $\theta_{23}$: 
\be
s_{23}^2>1/\bar{r}\simeq 0.4\,, \quad {\rm or} \quad
\theta_{23} \aprge 39^\circ\,.
\label{limit}
\ee
This is an independent confirmation of the conclusion that 
$\nu_\mu$-$\nu_\tau$ mixing should be rather large which follows from the 
data fits performed by the SK collaboration \cite{SuperK}. However, unlike
in the SK analysis, the lower bound (\ref{limit}) does not depend on the 
$\mu$-like data and is therefore complementary. If the excess of the 
sub-GeV e-like events is also at least partly due to the 
$\nu_e\leftrightarrow \nu_\mu$ oscillations, an even more stringent limit 
on $\theta_{23}$ would follow: $\theta_{23}\aprge 43^\circ$. 

2. With increasing statistics of the SK experiment and new
independent measurements of the primary cosmic ray flux it will 
become possible to give a definitive answer to the question of whether 
there is a non-vanishing up-down asymmetry of the multi-GeV electron 
events beyond the small asymmetry due to the geomagnetic effects. If the 
answer is positive, this would be a signature of non-zero 
mixing $\theta_{13}$. In a significant domain of the allowed values of 
$\Delta m_{32}^2$ the asymmetry is strongly enhanced by the resonance 
matter effects on neutrino oscillations and in particular, by the
parametric resonance. This makes it possible to probe even very small 
values of $\theta_{13}$ in the atmospheric neutrino experiments. 

The determination of the value of $\theta_{13}$ from the up-down 
asymmetry of the e-like events will require an independent knowledge 
of the values of $\Delta m_{32}^2$ and $\theta_{23}$. These could be
obtained, {\it e.g.}, from various samples of the $\mu$-like events 
which only weakly depend on $\theta_{13}$.

3. In principle, it is possible to experimentally disentangle the 
contributions from different resonance structures. 
Although this does not seem to be possible with the presently available 
data, such an analysis may become possible with future data with better
statistics and/or more accurate reconstruction of neutrino energies and 
directions. 

Clearly, selecting events in the non-vertical bins in which the trajectories 
that do not cross the earth's core dominate will allow one to estimate
the effects of the MSW resonance in the mantle. With high statistics 
e-like data this can be a realistic task. Clear identification of the 
MSW effect on oscillations of atmospheric neutrinos in the earth would 
be of paramount importance. 

For vertical bins one can use various energy cuts to discriminate 
between different resonance effects. 
Indeed, as we have pointed out in Sec. 3, the SK response 
function for multi-GeV events has a steep low-energy slope. Therefore 
using different energy cuts for leptons one can exclude the effects of the 
low energy peaks in the oscillation probability. For instance, if 
$\Delta m^2 \sim 2 \times 10^{-3}$ eV$^2$, then the response function 
corresponding to the threshold 1.33 GeV will cover all three resonances. 
However, with the threshold of 2 GeV, the effect of the parametric peak
can be significantly suppressed. 

4. It is interesting to note that if the preliminary BESS results 
\cite{BESS} are confirmed and the overall normalization of the atmospheric 
$\nu_e$ and $\nu_\mu$ fluxes is indeed somewhat below the current 
theoretical predictions, the SK data would imply a smaller deficiency of 
atmospheric $\nu_\mu$'s. 
At the same time, the data on e-like events as
well as the observed small value of the double ratio $R$ would 
therefore mean a significant excess of the e-like events. 
In such a situation the excess of e-like events due to the parametric effect 
can be enhanced. 

Indeed, a smaller deficiency of atmospheric $\nu_\mu$'s
means that the value of $\theta_{23}$, though still in a range of large
mixing, may be farther away from $45^\circ$. If $\theta_{23}$ is
noticeably {\em larger} than $45^\circ$, the excess of e-like events is
further enhanced by the  $(r s_{23}^2 - 1)$ factor [see (\ref{fluxe})].
Thus, in this case of the reduced flux normalization, the parametric
effects  would be especially important for understanding the SK atmospheric 
neutrino data. 

{\bf Note added :}

After this work  had been practically accomplished, the 
paper \cite{Fogli} has appeared in which the SK data were 
analyzed in the three neutrino oscillation scheme. The zenith angle
distributions are 
shown for large $\theta_{13}$, where parametric resonance effects play no
role. The  fit of the data in \cite{Fogli} agrees with our results: 
In fig.~16 of that paper, the allowed region in
the triangle for $\Delta m^2_{32} = 10^{-3}$ eV$^2$ 
corresponds to $s_{23}^2 > 0.5$ and 
small nonzero $\theta_{13}$.

\section*{Acknowledgments}

We would like to thank Eligio Lisi for useful discussions. 
One of us (PL) wishes to thank  Serguey Petcov for discussions on similar 
ideas  about three flavor oscillations in atmospheric
neutrinos. 

\appendix

\section*{Appendix : Cross-sections and event rates}

To describe the neutrino cross section we have
considered separately  the processes of
quasi-elastic scattering, single pion production
and multi--particle  production \cite{LLS}.
We have also included nuclear effects  according to  the treatment
of Smith and Moniz \cite{Smith-Moniz} - the 
nucleons  bound in the oxygen nucleus were assumed
to fill a Fermi sphere  up to a  maximum momentum
$p_F = 220$~MeV,  and to have a binding energy of 25~MeV.

The quasi elastic scattering   was  described following 
Llewellyn Smith \cite{Llewellyn},
using 
$F_A (Q^2) = -1.25\,(1+Q^2 / M_A^2 )^{-2}$
for the the axial-vector  form factor, 
with $M_A$ = 1.0 GeV \cite{Belikov}.
The nuclear effects  are important for  the quasi-elastic
cross section: the processes  where the final state nucleon
is scattered in  an occupied state
are prohibited  by the 
Pauli blocking  effect and the cross section 
is  reduced. The Fermi momentum  of the bound 
nucleons  also has the 
effect of  broadening the angular distribution of the final state
charged  leptons.

The cross section  for the single pion  production 
in the region $W < 1.4$~GeV  ($W$  is the mass of the   
hadronic system in 
the final state) was described following
Fogli and Nardulli \cite{Fogli_Nar}.
In this  region the most important  dynamical  effect is the presence 
of the $\Delta$ resonance.

All the other scattering  processes  were  described   using the standard
formula  for deep   inelastic scattering using the leading  order parton
distribution functions  (PDF's) of Gluck, Reya and Vogt \cite{GRV}.
In the monte carlo calculation we   used  the LUND algorithms
  \cite{LUND,JETSET}
to  construct physical  particles from the hadronic state
composed of the  scattered (anti-)quarks  and the nucleon  remnants
were described \cite{LEPTO} as  
a $qq$, $qqqq$ or $qqq\overline{q}$ system.

We have used a monte carlo method  for the calculation.
This  allows  to  include  all the dynamical  features  in detail,
including the important  nuclear effects, and also to
simulate  (at least  crudely) the experimental selection criteria
(in particular the `single-ring' and  containment  conditions).
Our  monte carlo also  generated  neutral--current  events,  but we 
have not  considered
the possibility of the mis-classification of  NC events as
CC events.

To  simulate single-ring events  we 
selected  the events with a charged  lepton in the appropriate  range of
momentum, and  required  in addition the 
absence of  photons  or  additional  charged  particles 
above the Cherenkov threshold.  
The  single ring requirement is important  because it
preferentially selects  lower  energy neutrinos  and therefore
changes the response  function for the   different  categories of 
events.
For the containment  requirement,   we assumed that all electron events
in the fiducial  volume  were contained;  for each muon event
generated  a  neutrino interaction point in 
the fiducial   volume   and checked whether  the   range 
in water of the final state $\mu^\pm$   was 
shorter  or  longer  than the distance along the trajectory
from the  interaction point to the PMT surface.

Our no--oscillation calculation   is  
approximately  20\% lower in normalization
than the Super--Kamiokande  monte carlo
for all  five categories of events (e-like and $\mu$--like
sub-GeV and multi-GeV  fully contained  events  and partially
contained  events),  with very good agreement in 
the zenith angle  distributions.
The absolute  normalization of the  calculation  is  sensitive 
to details  such as the minimum amount of Cherenkov light
that a  charged particle needs to have in order to produce
an additional  visible ring. For the purposes  of  our discussion we  
find the agreement to be good.

\newpage
\centerline{\large Figure captions}

\vglue 0.4cm
\noindent
\noindent 
Fig. 1. Dependence of the transition probability $P_2$
on neutrino energy for $\Delta m_{32}^2=2\times 10^{-3}$ 
eV$^2$  (a) 
for $\sin^2 2\theta_{13} = 0.025$
and different values 
of  $\cos \Theta_{\nu}: - 0.98$ (solid  curve),
$-0.88$ (long-dashed curve),  and
$-0.85$ (short-dashed curve);
(b) for $\cos \Theta_{\nu} = - 0.98$
and  different values of $\sin^2 2\theta_{13}$: 
$0.014$ (short-dashed curve),
$0.025$ (long-dashed curve),  $0.057$ (solid curve). 
Neutrino energy in eV. 

\noindent
Fig.~2. Zenith angle dependences of multi-GeV events. 
(a) e-like events: solid line -- no oscillations,  
dashed line corresponds to oscillations with  
$\sin^2 2\theta_{13} = 0.06$,
$\sin^2 \theta_{23} = 0.75$ and 
$\Delta m^2_{32} = 10^{-3}$ eV$^2$.
(b) $\mu$-like events:  
upper solid histogram is for  no-oscillations case,  
the dashed histogram  -- oscillations 
with $\sin^2 2\theta_{13} = 0.06$,
$\sin^2 \theta_{23} = 0.75$ and
$\Delta m^2_{32} = 10^{-3}$ eV$^2$, 
the lower solid histogram  -- two-neutrino oscillations 
with the same parameters but $\sin^2 2\theta_{13} = 0$. 

\noindent
Fig.~3.   
Dependence of the up-down asymmetries of 
multi-GeV (a) e-like  events,  
$A^{U/D}_e(0.6,1)$,  
and (b) $\mu$-like  events, $A^{U/D}_{\mu}(0.6,1)$, on 
$\Delta m^2_{32}$ for  
$s^2_{23} = 0.75$ and  different values of $\sin^2 2\theta_{13}$:   
$\sin^2 2\theta_{13} = 0.03$ (dashed curve),   
$0.06$ (solid curve), $0.10$ (dot - dashed) curve. 
The dotted curve in fig.~3b shows the asymmetry for 
pure $\nu_{\mu} \leftrightarrow \nu_{\tau}$ oscillations   
with $s^2_{23}= 0.75$. 
(c). The same as in fig.~3a but for different bins: 
$A^{U/D}_e(0.84,1)$ (dashed curve), $A^{U/D}_e(0.60,0.84)$ (dot-dashed 
curve) and $A^{U/D}_e(0.60,1)$ (solid curve); $\sin^2 2\theta_{13}=0.06$. 

\noindent
Fig.~4.    Dependence of the up-down asymmetries of multi-GeV 
(a) e-like events,  $A^{U/D}_e(0.6,1)$, 
and (b) $\mu$-like events,  $A^{U/D}_{\mu}(0.6,1)$, 
on $\sin^2 2\theta_{13}$ 
for  $s^2_{23} = 0.75$ and  different values of $\Delta m^2_{32}$:  
$0.5 \times 10^{-3}$  eV$^2$ (dot-dashed curve), 
$1.7 \times 10^{-3}$  eV$^2$ (dashed curve), 
$3 \times 10^{-3}$  eV$^2$ (solid curve).  
The squares on the curves represent the CHOOZ bound: parts of the
curves on the right of the squares are excluded. 

\noindent
Fig.~5.  
Dependence of the up-down asymmetries 
(a) of e-like events, $A^{U/D}_e(0.6,1)$, 
and (b) of $\mu$-like events, $A^{U/D}_{\mu}(0.6,1)$, on
$s^2_{23}$ for $\Delta m^2_{32} = 1.7\times 10^{-3}$  eV$^2$
and  different values of $\sin^2 2\theta_{13}$:
$\sin^2 2\theta_{13} = 0.03$ (dashed curve), 
$0.06$ (solid curve), $0.10$ (dot - dashed curve).
The dotted curve in fig.~5b shows the asymmetry for
pure $\nu_{\mu} \leftrightarrow \nu_{\tau}$ oscillations
with $s^2_{23}= 0.75$.

\noindent
Fig.~6. Zenith angle dependence of the multi-GeV  e-like
events. The solid  histogram  is for the no-oscillation case. 
The dashed  histogram is calculated for 
$\Delta m^2 = 1.7\times 10^{-3}$~eV$^2$,
$\sin^2 2 \theta_{13}  = 0.10$
and $\sin^2 2 \theta_{23}  = 0.75$.
The points are the 535 days data  of
the Super-Kamiokande. 

\noindent
Fig.~7. Iso-asymmetry contour plot for multi-GeV e-like events in the 
$(\sin^2 2\theta_{13}, \Delta m_{32}^2)$ plane for $s_{23}^2=0.75$. 
The closed curve corresponds to 
$A_e^{U/D}=0.264$. The other curves (from bottom upward): 
$A_e^{U/D}=0.15$, 0.175, 0.20, 0.225, 0.25, 0.275, 0.30, 0.325, 0.35, 
0.375, 0.40, 0.425 and 0.45. The shaded area shows the region 
excluded by CHOOZ.

\noindent
Fig.~8.  Zenith angle distribution of the e-like events 
in  the sub-GeV range. Solid histogram -- without oscillations; 
dashed histogram --  oscillations with  
$\Delta m^2_{32} = 0.3 \times 10^{-3}$ eV$^2$, 
$\sin^2 2\theta_{13} = 0.1$ and 
$\sin^2 \theta_{23} = 0.75$.
The points are the 535 days data  of
the Super-Kamiokande. 

\noindent
Fig.~9.  
(a) The ratio 
of the e-like events rates
in the sub-GeV region with and without oscillations  
as the function of  $\Delta m^2_{32}$ for 
$\sin^2 \theta_{23} = 0.75$ (solid curves) and   
$\sin^2 \theta_{23} = 0.65$ (dashed curves) 
and different values of $\sin^2 2\theta_{13}$. From the lowest 
to the highest curve: $\sin^2 2\theta_{13} = 0.03$, 
0.06, 0.10, 0.20.  
(b) the same as in fig.~7a but  for $\mu$-like events. $\sin^2 2
\theta_{13} = 0.03, 0.06, 
0.10, 0.20$  (from  the highest
to the lowest  curve). The dotted curve corresponds to pure 
$\nu_{\mu} \leftrightarrow \nu_{\tau}$ 
oscillations with $\sin^2 \theta_{23} = 0.5$.  

\newpage

\begin{figure} [t]
\centerline{\psfig{figure=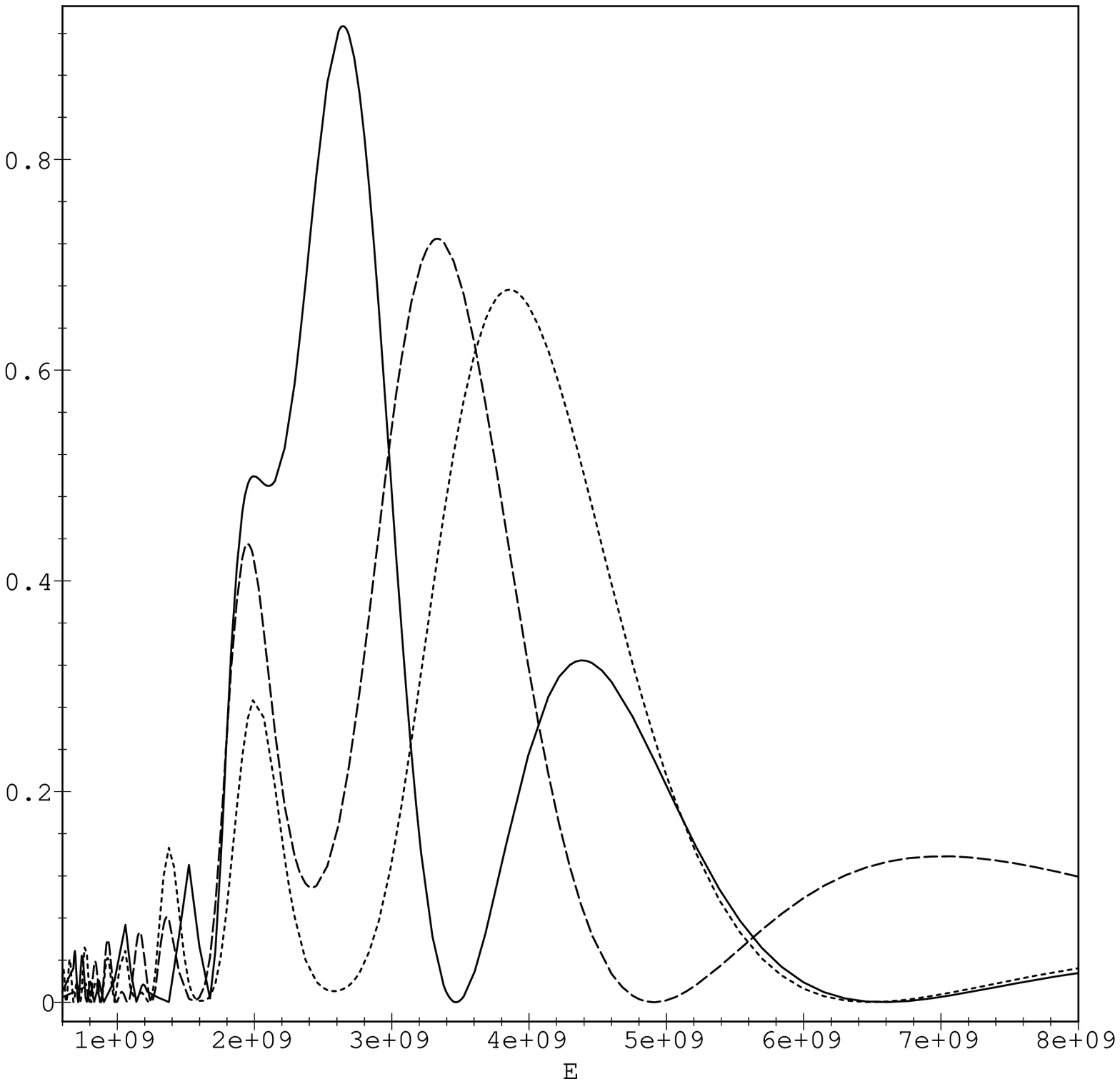,height=21cm}}
\centerline{\large \tt Figure 1a}
\end{figure}

\begin{figure} [t]
\centerline{\psfig{figure=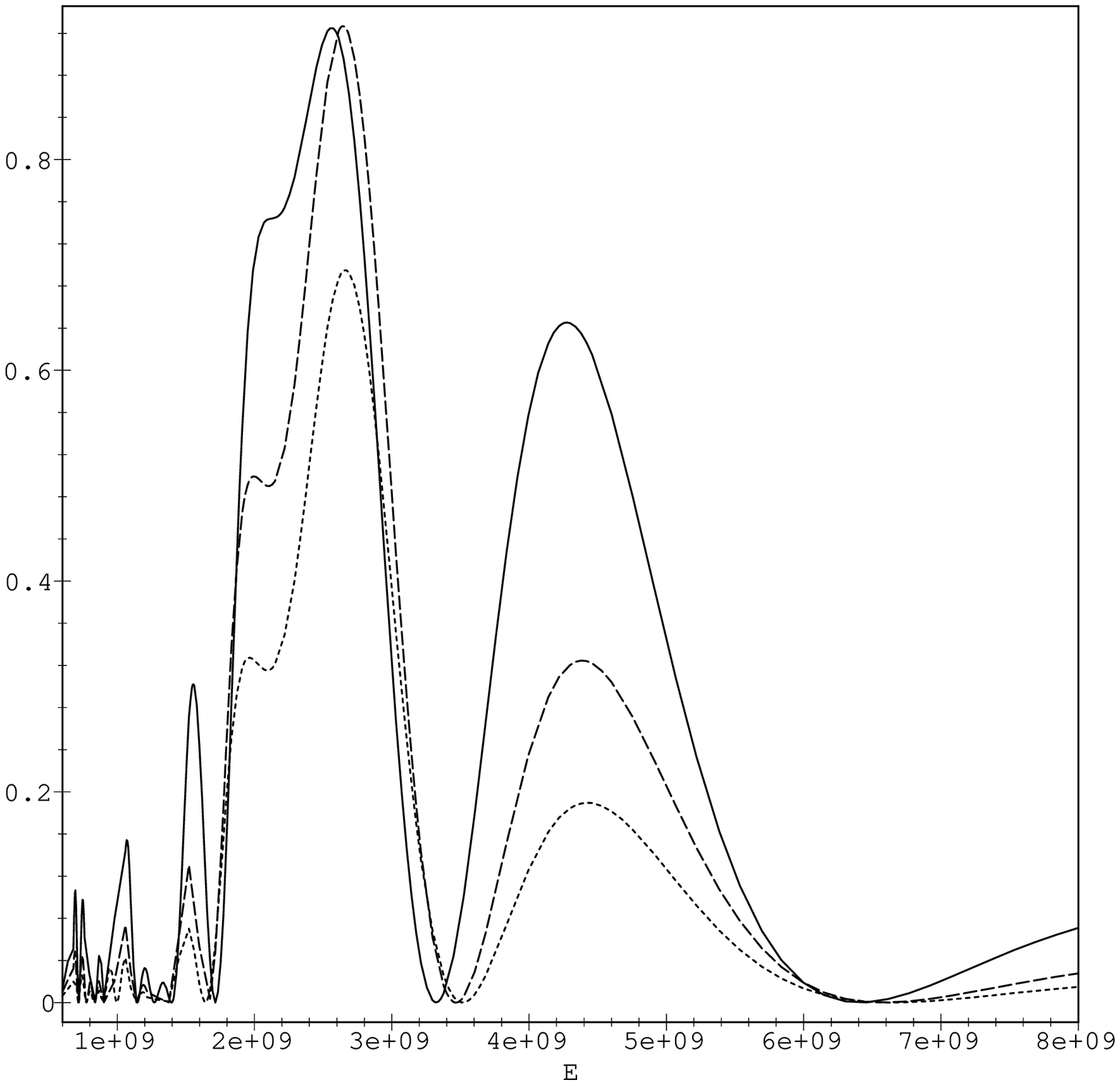,height=21cm}}
\centerline{\large \tt Figure 1b}
\end{figure}

\begin{figure} [t]
\centerline{\psfig{figure=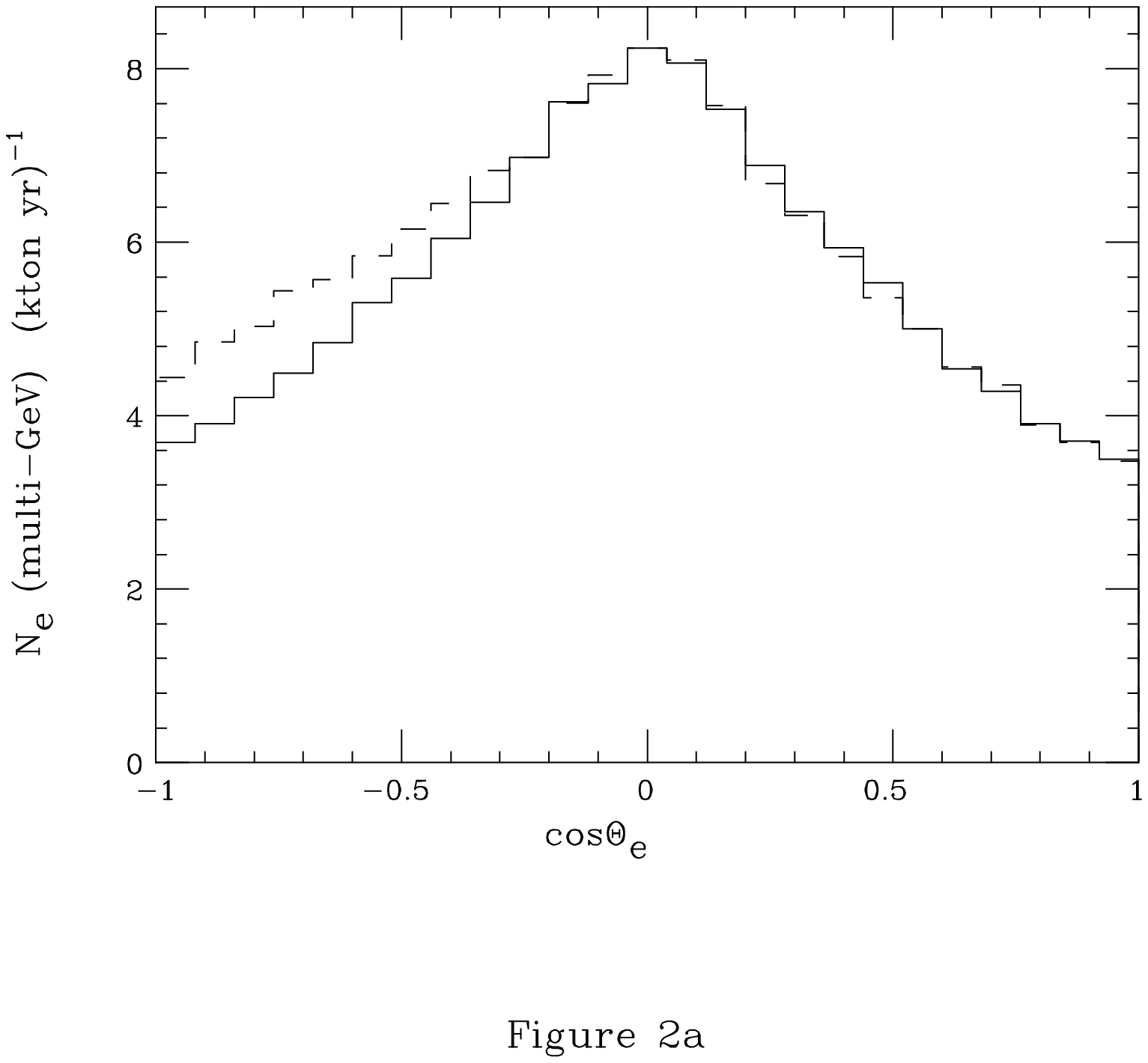,height=21cm}}
\end{figure}

\begin{figure} [t]
\centerline{\psfig{figure=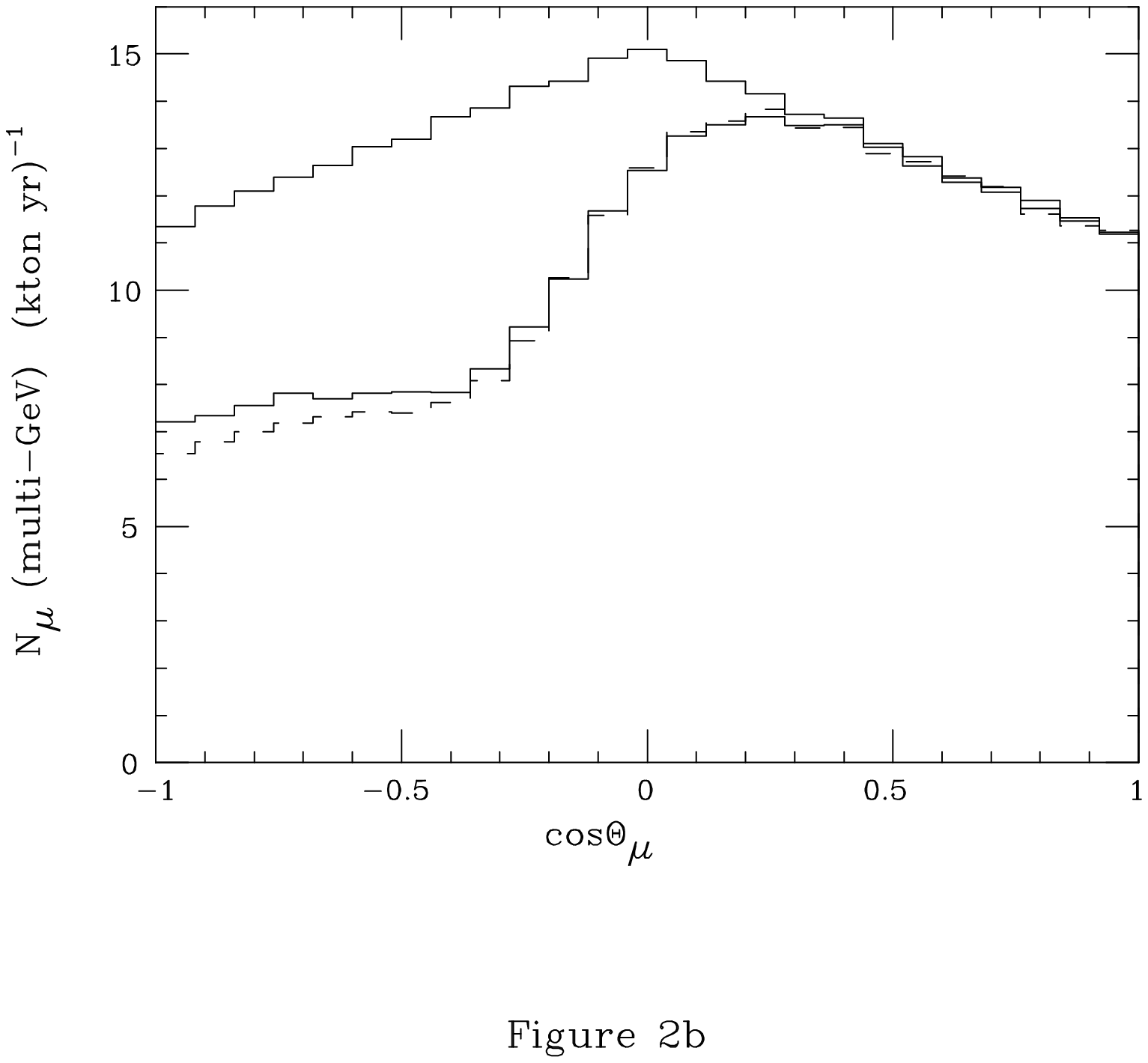,height=21cm}}
\end{figure}

\begin{figure} [t]
\centerline{\psfig{figure=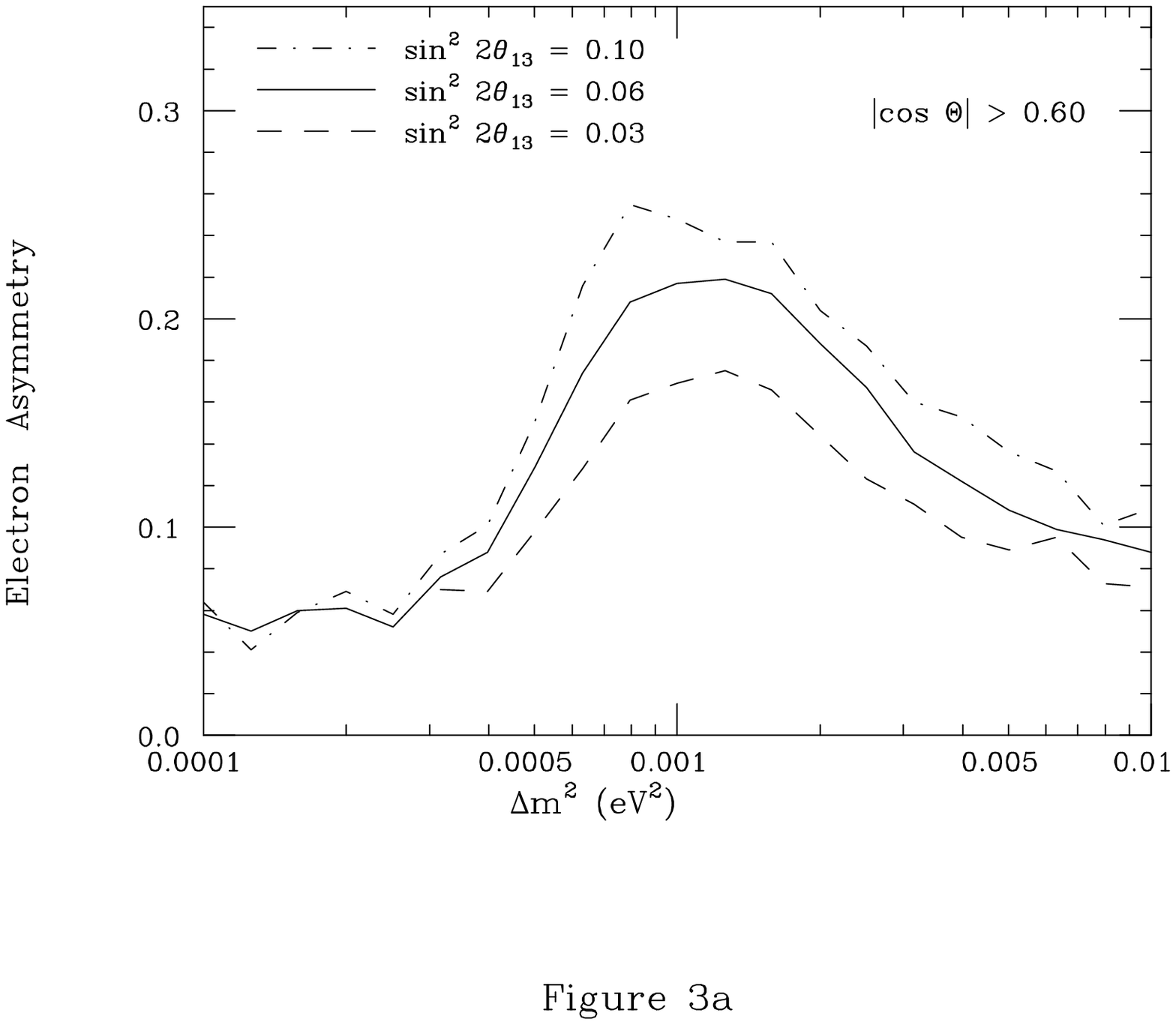,height=21cm}}
\end{figure}

\begin{figure} [t]
\centerline{\psfig{figure=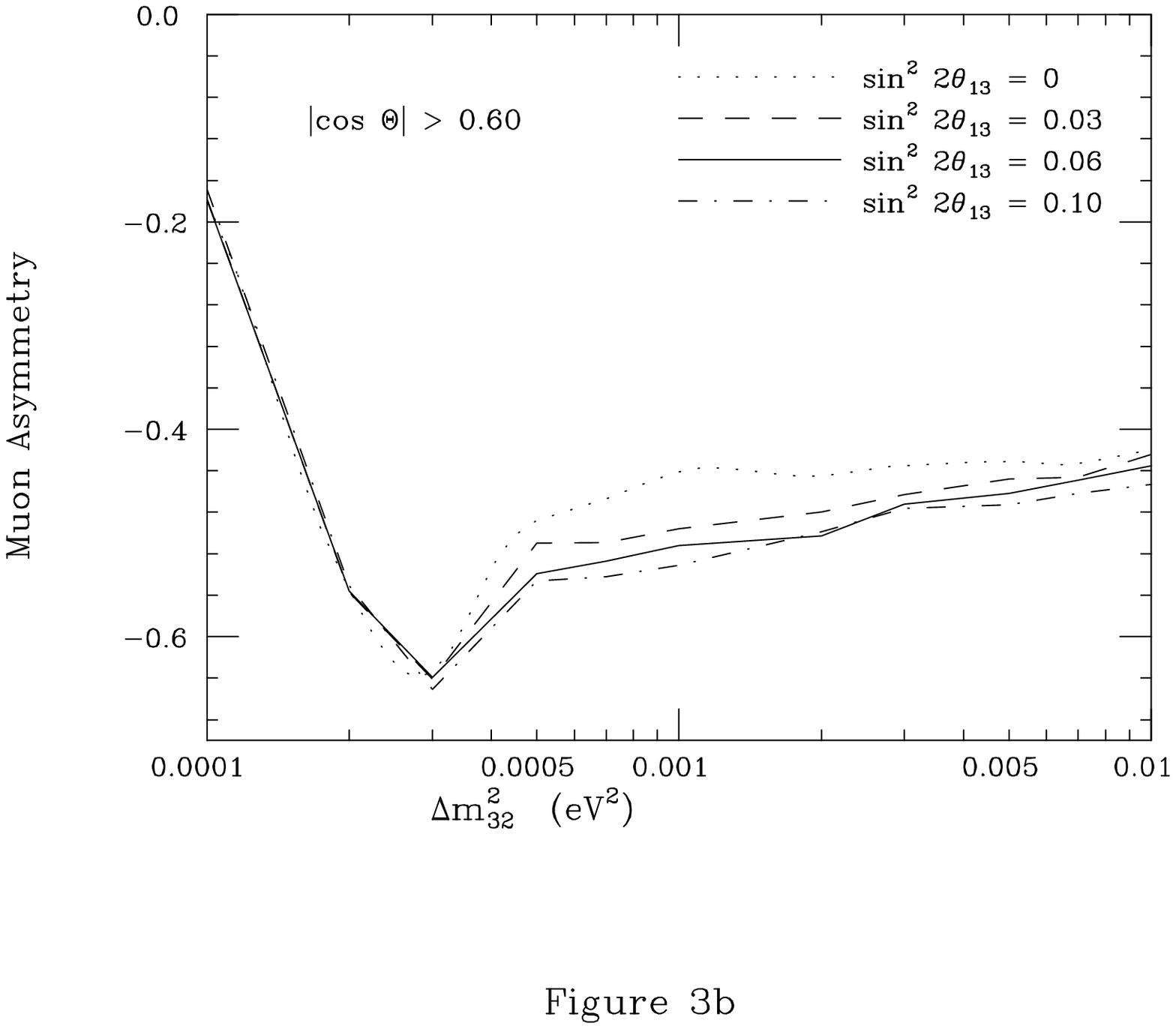,height=21cm}}
\end{figure}

\begin{figure} [t]
\centerline{\psfig{figure=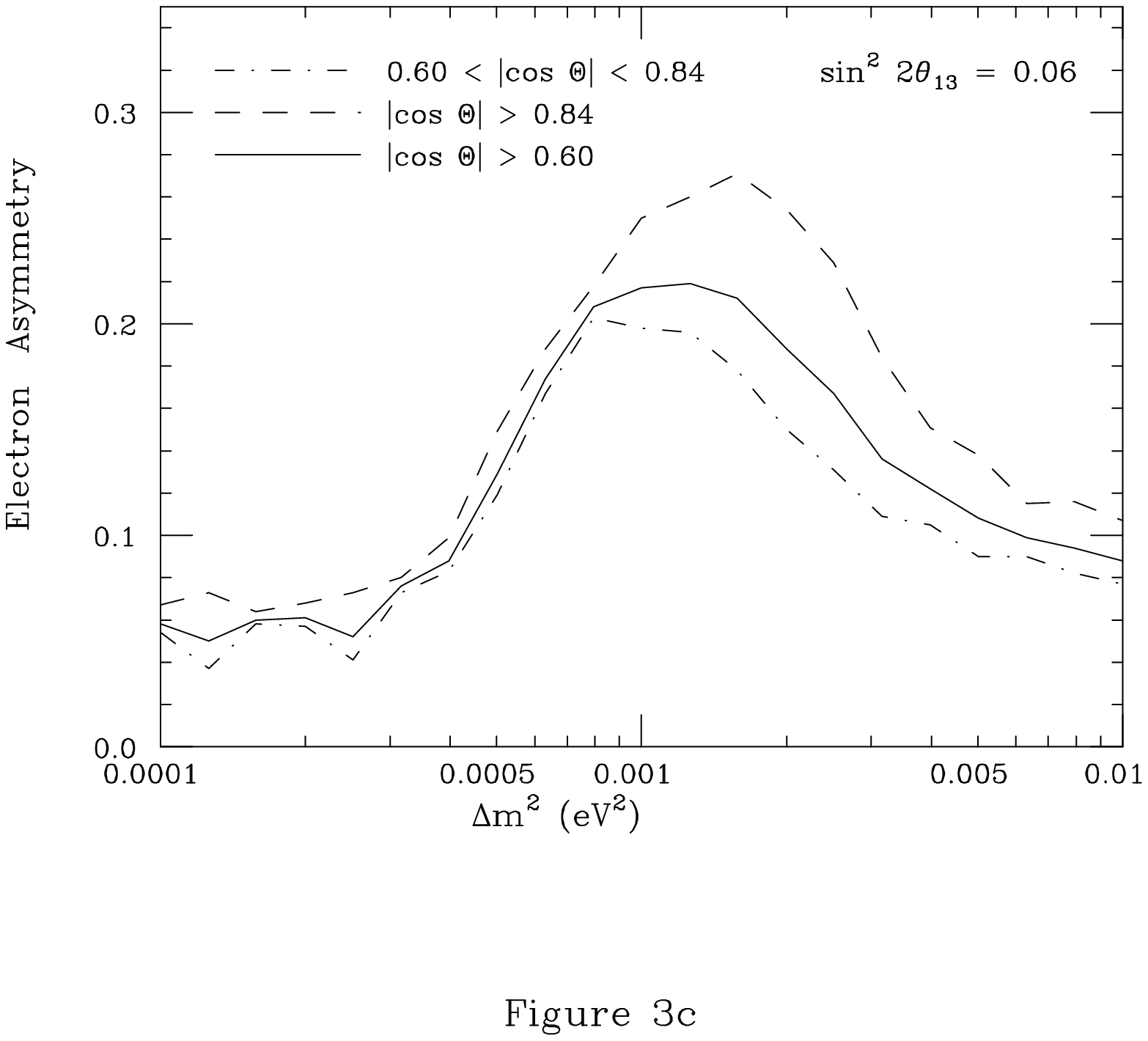,height=21cm}}
\end{figure}

\begin{figure} [t]
\centerline{\psfig{figure=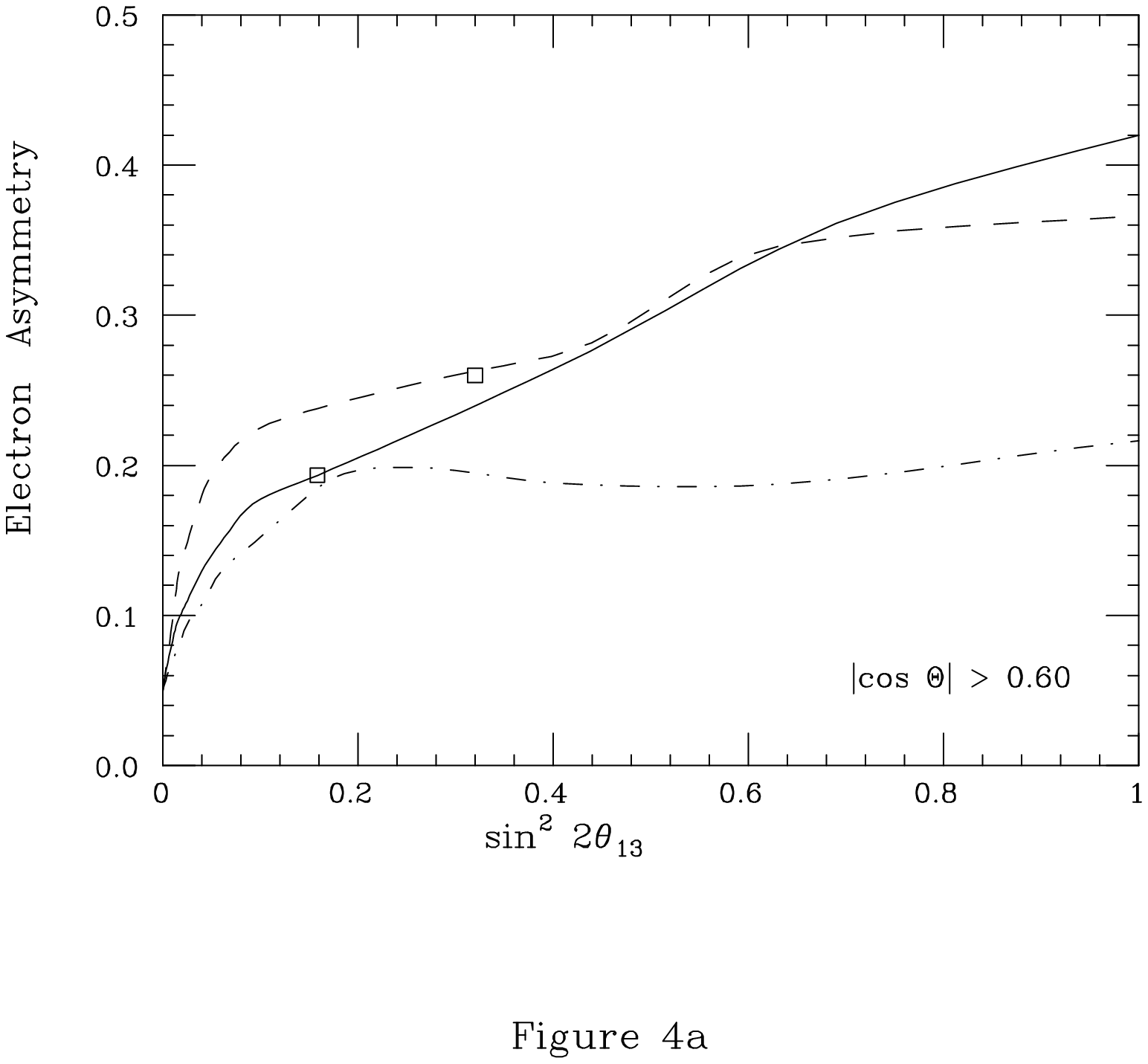,height=21cm}}
\end{figure}

\begin{figure} [t]
\centerline{\psfig{figure=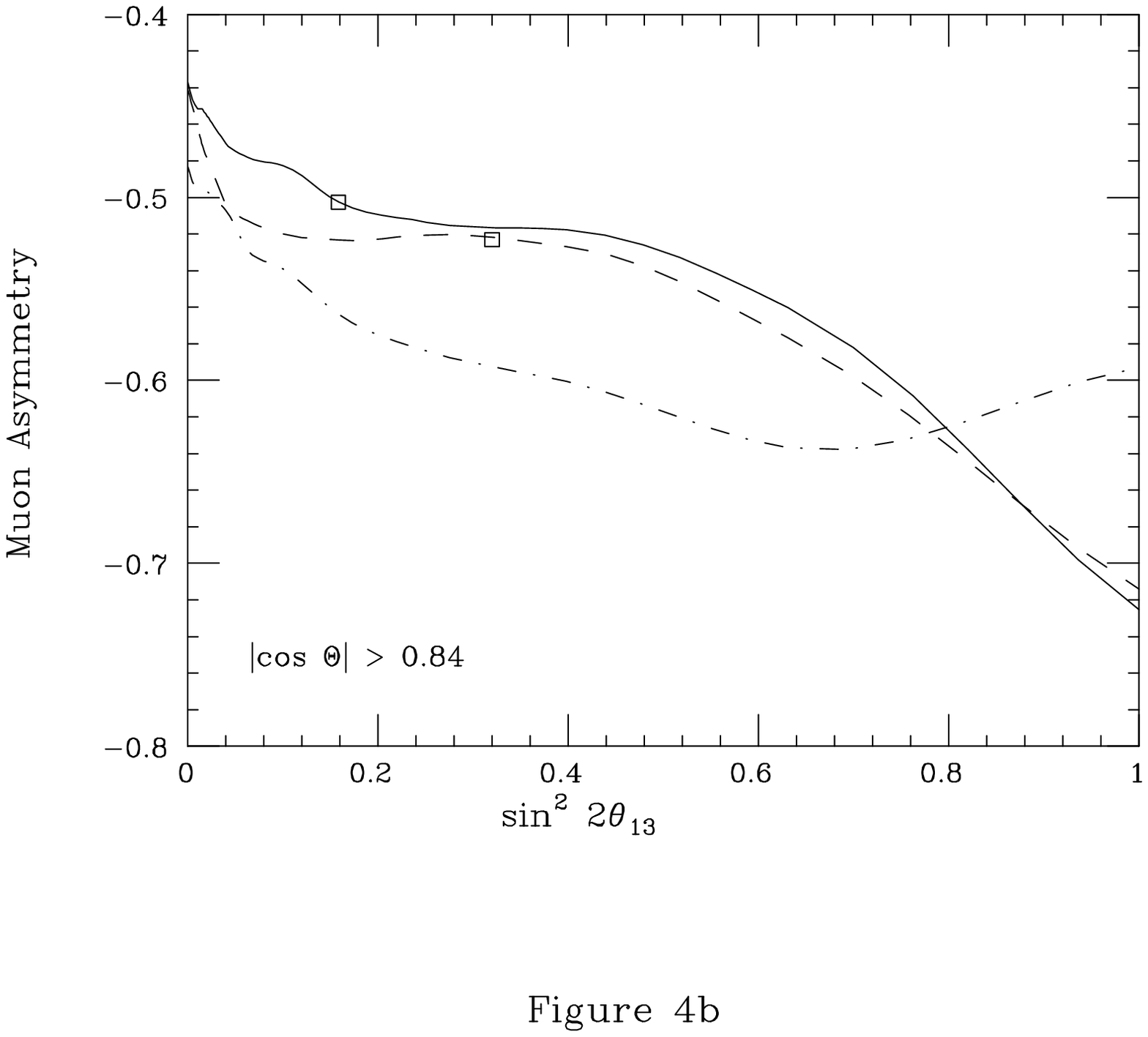,height=21cm}}
\end{figure}

\begin{figure} [t]
\centerline{\psfig{figure=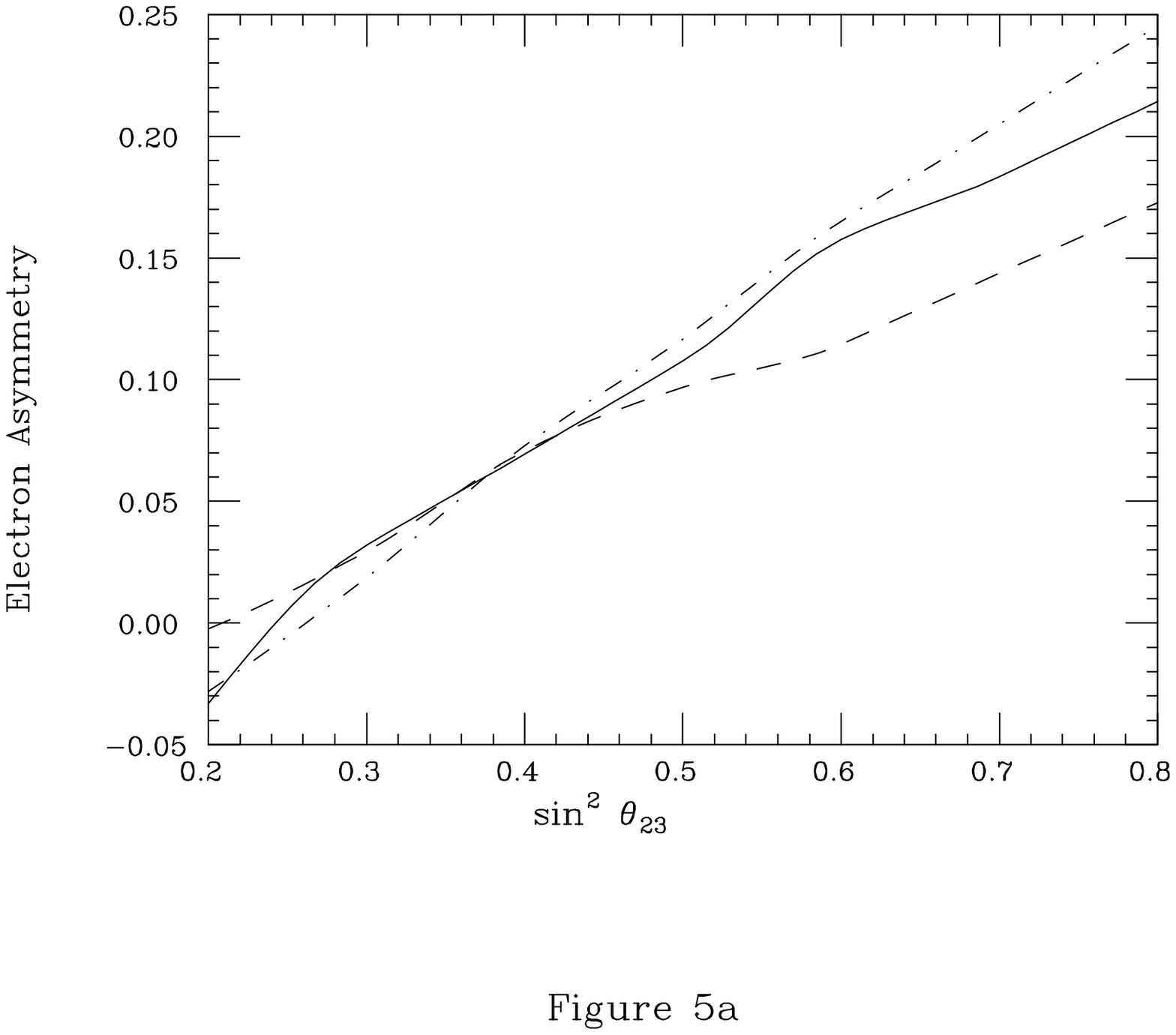,height=21cm}}
\end{figure}

\begin{figure} [t]
\centerline{\psfig{figure=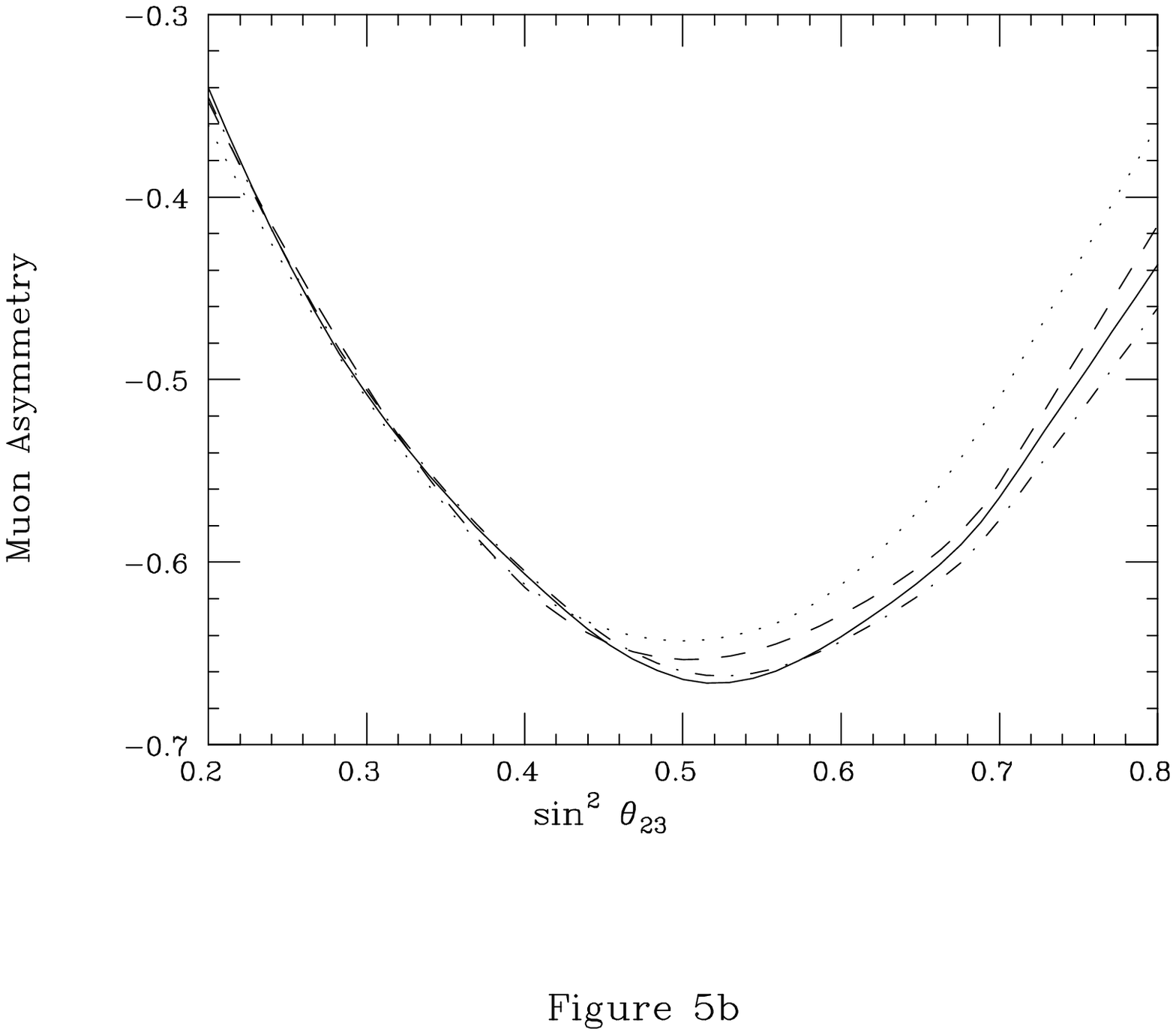,height=21cm}}
\end{figure}

\begin{figure} [t]
\centerline{\psfig{figure=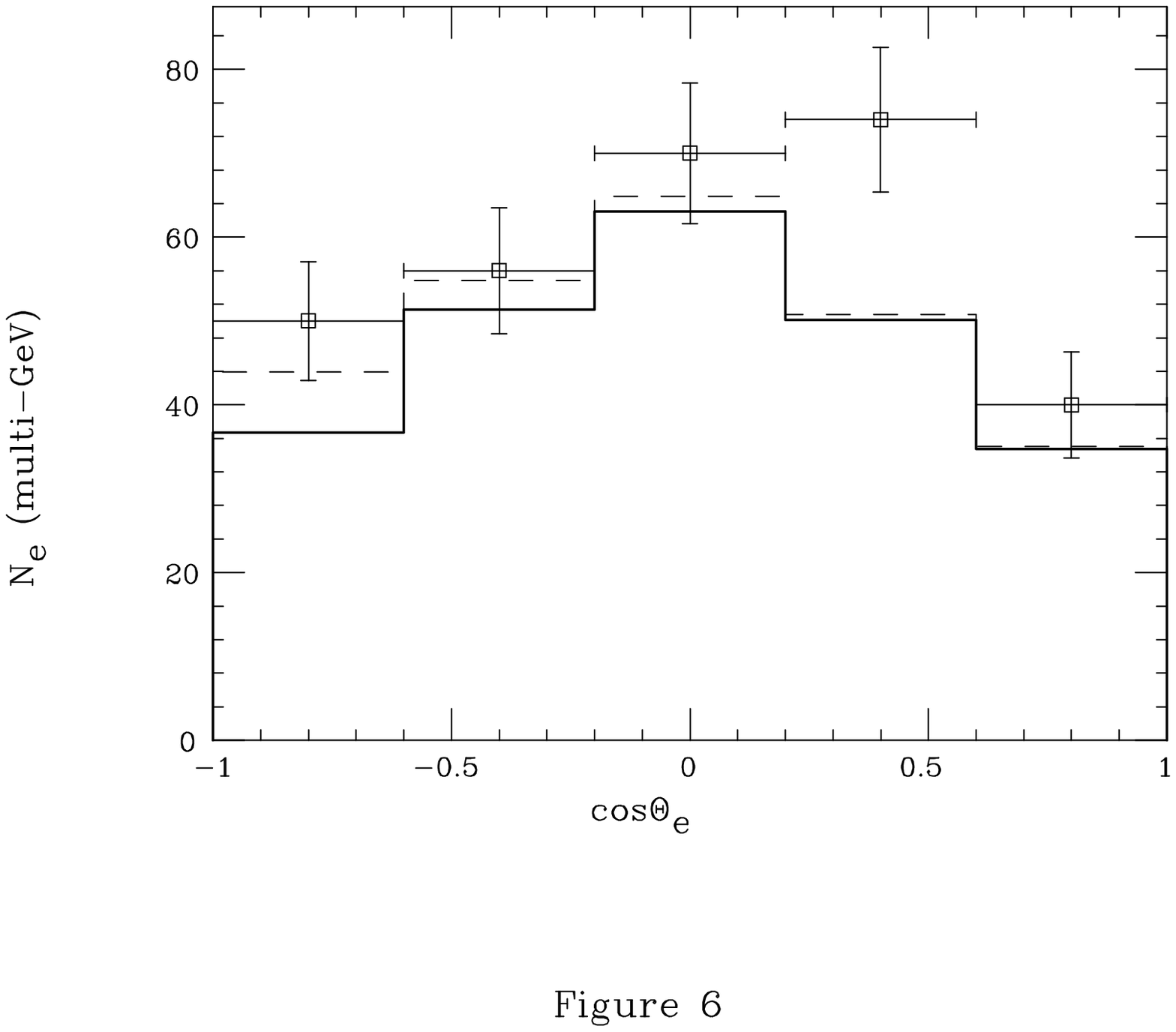,height=21cm}}
\end{figure}

\begin{figure} [t]
\centerline{\psfig{figure=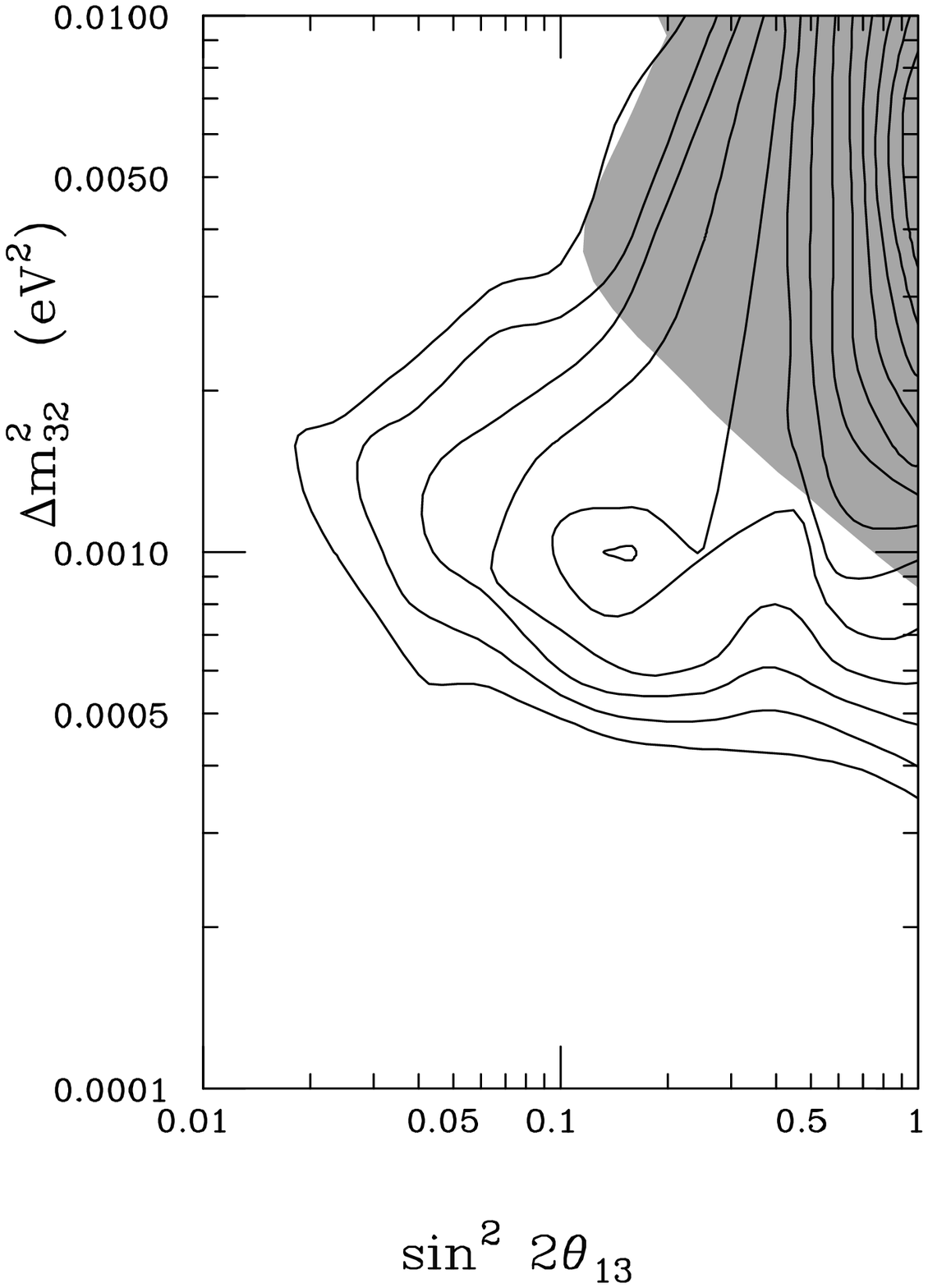,height=21cm}}
\centerline {\bf \Large {Figure 7} }
\end{figure}

\begin{figure} [t]
\centerline{\psfig{figure=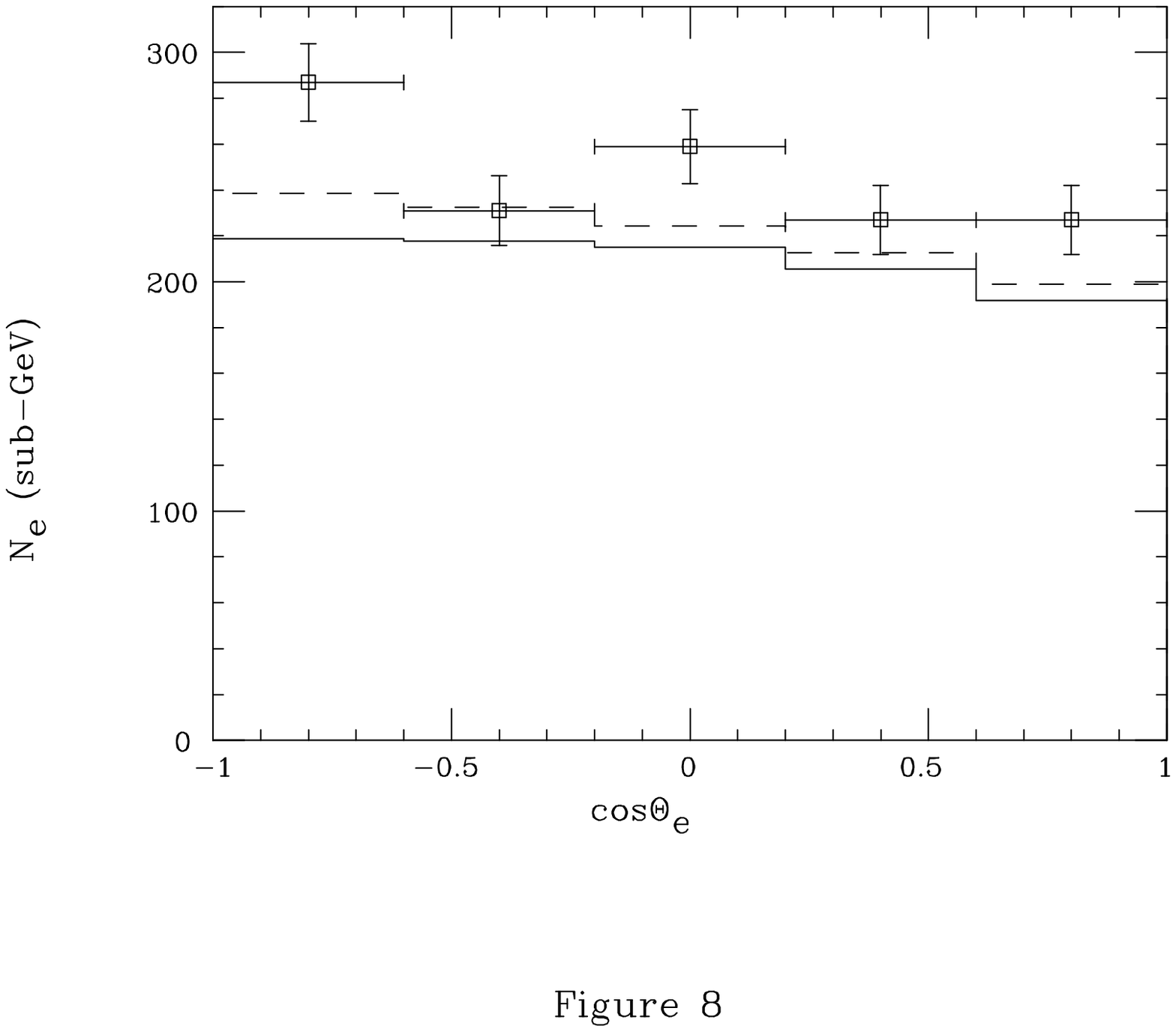,height=21cm}}
\end{figure}

\begin{figure} [t]
\centerline{\psfig{figure=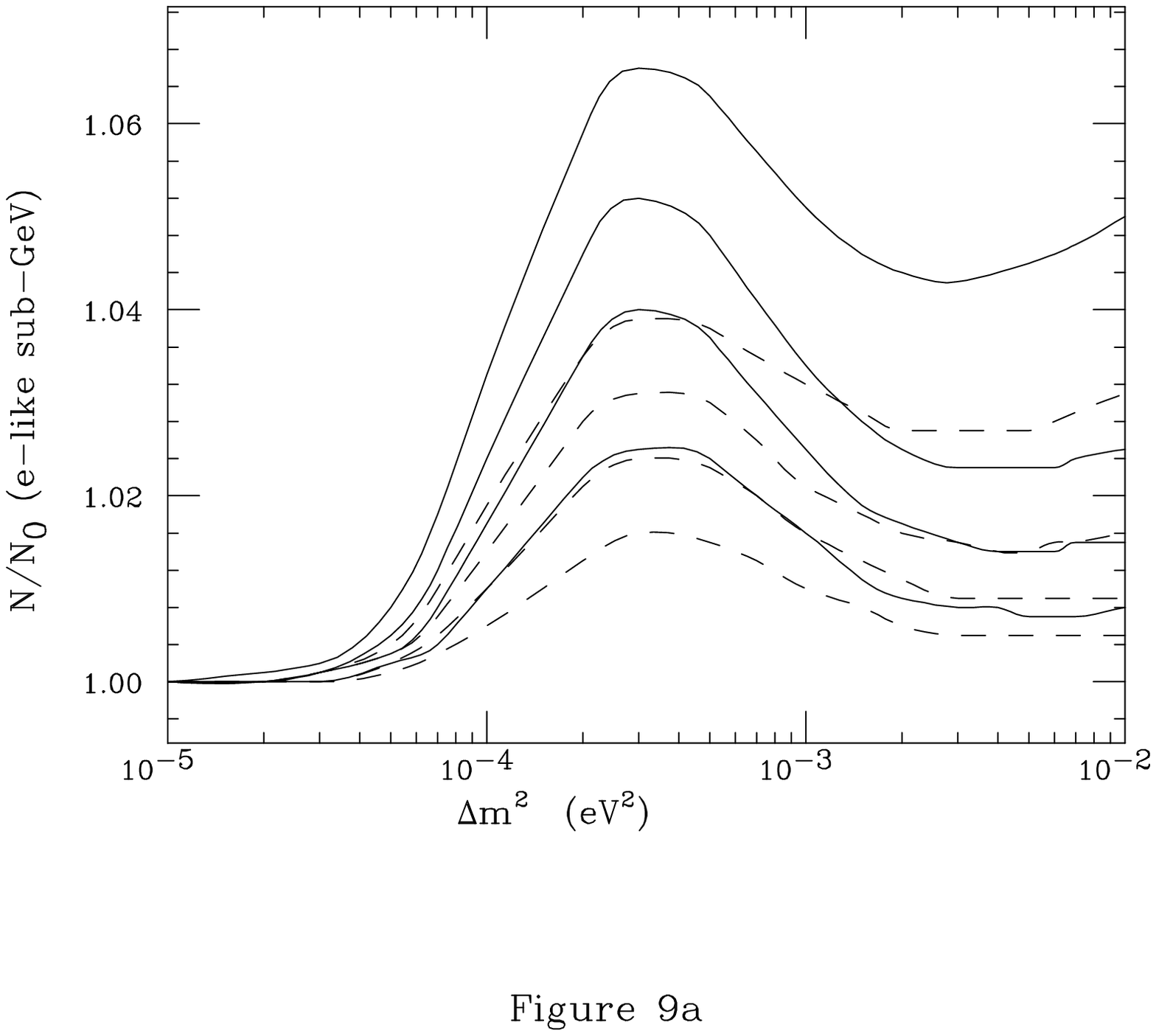,height=21cm}}
\end{figure}

\begin{figure} [t]
\centerline{\psfig{figure=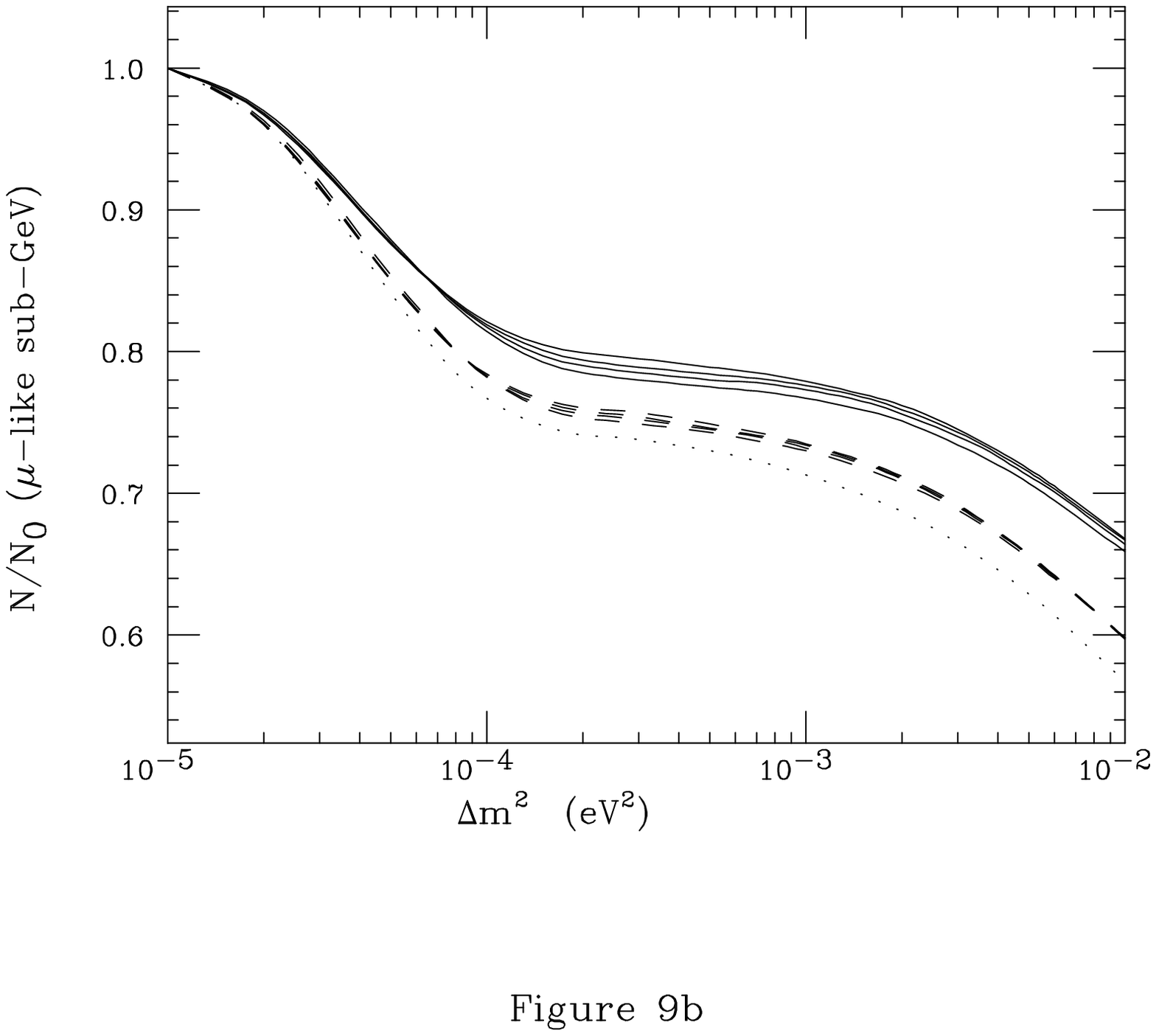,height=21cm}}
\end{figure}

\end{document}